\documentclass[10pt,aps,prb,twocolumn,floatfix,floats,showpacs,superscriptaddress,raggedbottom]{revtex4-1}
\usepackage{graphicx,latexsym}
\usepackage{dcolumn}
\usepackage{amssymb,amsmath,bm}
\pdfoutput=1

\begin{document}


\title{Spin and impurity effects on flux-periodic oscillations in core-shell nanowires}

\author{Tomas Orn Rosdahl}
\email{torosdahl@gmail.com}
\affiliation{Science Institute, University of Iceland, Dunhaga 3,
         IS-107 Reykjavik, Iceland}
\author{Andrei Manolescu}
\affiliation{School of Science and Engineering, Reykjavik University, Menntavegur 1, IS-101 Reykjavik, Iceland}
\author{Vidar Gudmundsson}
\affiliation{Science Institute, University of Iceland, Dunhaga 3,
         IS-107 Reykjavik, Iceland}

\begin{abstract}
We study the quantum mechanical states of electrons situated on a cylindrical
surface of finite axial length to model a semiconductor core-shell nanowire.
We calculate the conductance in the presence of a longitudinal magnetic field
by weakly coupling the cylinder to semi-infinite leads. Spin effects are accounted for through
Zeeman coupling and Rashba spin-orbit interaction (SOI). Emphasis is on manifestations of
flux-periodic (FP) oscillations and we show how factors such as impurities, contact geometry
and spin affect them. Oscillations survive and remain periodic in the presence
of impurities, noncircular contacts and SOI, while Zeeman splitting results
in aperiodicity, beating patterns and additional background fluctuations.
Our results are in qualitative agreement with recent magnetotransport experiments
performed on GaAs/InAs core-shell nanowires. Lastly, we propose methods of data
analysis for detecting the presence of Rashba SOI in core-shell systems
and for estimating the electron g-factor in the shell.
\end{abstract}


\maketitle

\section{Introduction}

Recent years have seen significant progress in the development
and fabrication of semiconductor nanostructures. Nanowires
of diameters of the order of $10-100$ nm can now be
grown.\cite{Bakkers2003,Li2006,Thelander2006,Yang2010} Core-shell
nanowires are composed of a thin layer (shell) surrounding a core in a
tubular geometry. The cross section may be circular, but also polygonal
(e.\ g.\ hexagonal) reflecting the lattice structure of the materials.
If both shell and core are semiconductors, they may be chosen such that
the difference in conduction band energies forms a potential barrier
confining carriers to either the core\cite{vanTilburg2010,Popovitz2011}
or the shell.\cite{Rieger2012,Jung2008} Recent examples include nanowires
composed of an InAs shell grown on a GaAs core resulting in the formation
of a conductive electron gas in the shell which may be further augmented
by modulation doping the core.\cite{Gul2014,Blomers2013} Such systems
provide a fascinating means for studying fundamental quantum effects
such as interference.

A prominent interference phenomenon due to the interaction of
electromagnetic potentials with charged particles is the Aharonov-Bohm
effect,\cite{Aharonov1959} which arises because wave functions that enclose
a magnetic flux acquire a flux-dependent phase shift. Furthermore,
the corresponding energy levels are periodic in the flux\cite{Byers1961}
and FP oscillations have indeed been observed for example in the
resistance\cite{Webb1985} and magnetization\cite{Levy1990} of rings pierced
by a magnetic flux. A couple of theoretical papers have addressed FP
oscillations in core-shell systems and found that spin ruins the
periodicity due to Zeeman splitting. In Ref.\ \onlinecite{Gladilin2013} the magnetization
of closed shells was analyzed and the robustness of FP oscillations to a nonzero
shell thickness demonstrated. Additionally, it was shown that while static donor impurities affect
the shape and phase of magnetization oscillations strongly, electron-electron
interaction tends to weaken their effects on the oscillations. Magnetoconductance oscillations in cylindrical
quasi-one-dimensional shells were treated in Ref.\ \onlinecite{Tserkovnyak2006}
and FP oscillations predicted assuming a narrow surface confinement.
Later, magnetoresistance measurements on core-shell nanowires
revealed the existence of FP oscillations,\cite{Jung2008}
which were recently shown to manifest because transport is mediated by closed-loop
angular momentum states encircling the core, tying the oscillations
to the spectrum.\cite{Gul2014}

In this paper, we analyze the energy spectrum, charge and current densities
of electrons confined to a closed cylindrical surface of finite length pierced
by a longitudinal magnetic flux. We calculate the magnetoconductance of
the finite system by coupling it to leads. In particular, we focus on
FP oscillations in both conductance and spectrum and discuss
the effects of core donor impurities and electron spin,
which is included through Zeeman splitting and Rashba SOI and may thus affect
transport nontrivially. The experimentally-relevant effects of nonuniform coupling to leads
are also discussed. While donor impurities dampen conductance oscillations,
they remain resolvable after extensive averaging over multiple random impurity
configurations, assuming realistic donor densities. Using parameters comparable
to those reported in Ref.\ \onlinecite{Gul2014}, we attribute background conductance
oscillations, which are superimposed on the FP oscillations, to an interplay between
the finite system length and Zeeman splitting,
propose means for detecting the presence of Rashba SOI and discuss a method for
estimating the shell electron g-factor based on transport data.

In Sec.\ \ref{section:Theory} we describe the closed-system model and transport
formalism and discuss results in Sec.\ \ref{section:ResultsNoImp}. The case
with impurities is treated in Sec.\ \ref{section:Impurities}
and in Sec.\ \ref{section:ExpComp} we give a comparison with recent experimental data.
Finally, we offer concluding remarks in Sec.\ \ref{section:Conclusions}.

\section{Quantum mechanical model for core-shell nanowires} \label{section:Theory}

We consider a cylindrical core-shell nanowire of radius $r_0$ and length
$L_0$ where the shell and core are composed of different semiconductors
such that the difference in conduction band energies confines conduction
electrons to the shell with negligible wavefunction leakage into
the core. We assume that the shell thickness is small compared to
$r_0$ and $L_0$ such that only the lowest radial mode is occupied and
approximate the shell as infinitely thin. In principle
$r_0$ corresponds to the mean radius of the shell and thus we model the
nanowire as a two-dimensional cylindrical surface.

\subsection{Closed wire}

The starting point for a quantum mechanical model of electrons confined
to a cylindrical surface of finite length, described by the cylindrical 
coordinates $\boldsymbol{r} = (r_0,\varphi,z)$, is the single-electron
Hamiltonian. In the presence of a longitudinal magnetic field
$\boldsymbol{B} = B\boldsymbol{\hat{z}}$ with vector potential $\boldsymbol{A} =
\frac{1}{2}Br_0\boldsymbol{\hat{\varphi}}$, the kinetic momentum $\boldsymbol{p} + e\boldsymbol{A} = 
-i\hbar\boldsymbol{\nabla} + e\boldsymbol{A}$ yields the kinetic energy term
\begin{equation}
H_O = \frac{\hbar^2}{2m_e}\left[ \frac{1}{r_0^2}\left( \frac{\partial_{\varphi}}{i} + 
\frac{\Phi}{\Phi_0} \right)^2 + \left( \frac{\partial_z}{i} \right)^2  \right] \label{eq:Kin}.
\end{equation}
Here, $m_e$ is the effective mass of shell conduction electrons, $\Phi = r_0^2 B\pi$
the longitudinal magnetic flux piercing the cylinder and $\Phi_0 = h/e$
the magnetic flux quantum. To model finite cylinder length we use 
hard-wall potentials at the cylinder edges,
\begin{equation}
  V_c(z) = \left\{ 
  \begin{array}{l l}
    0 & \quad \text{if $0 < z < L_0$}\\
    \infty & \quad \text{otherwise}. \\
  \end{array} \right. \label{eq:conf}
\end{equation}

The electron spin yields a Zeeman term which in a longitudinal
field has the form
\begin{equation} H_Z = \frac{\hbar \omega_c g_e m_e}{4m_0} \sigma_z 
\label{eq:Zeeman}
\end{equation}
where $m_0$ is the free electron mass,
$g_e$ the effective g-factor of shell conduction electrons
and $\omega_c = eB/m_e$ the cyclotron frequency. In addition, we model Rashba
SOI in core-shell geometries as arising due to interactions between
the shell electron gas and ionized donors in the core, producing an approximately
radial electric field. For our model it reads\cite{Winkler,Bringer2011,Mehdiyev2009}
\begin{equation}
 H_{SOI} = \frac{\alpha}{\hbar} \left[\sigma_{\varphi}p_z - \sigma_z \left(p_{\varphi} + 
eA_{\varphi} \right) \right], \label{eq:Rashba}
\end{equation}
where $\alpha$ determines the SOI strength and $\sigma_{\varphi} = \cos{(\varphi)}\sigma_y
- \sin{(\varphi)}\sigma_x$ describes the tangential spin projection. A Dresselhaus type SOI,
arising due to inversion asymmetry in the crystal structure of the
shell, may also be included.  Such effects are typically minor in
InAs compared to those of the Rashba term, and we have therefore
chosen to neglect them in the present context.

The total single-electron Hamiltonian of the closed central system $H_S$ is the sum of the terms 
in Eqs.\ (\ref{eq:Kin}) to (\ref{eq:Rashba}),
\begin{equation} 
H_S = H_O + V_c + H_Z + H_{SOI}. 
\label{eq:HS} 
\end{equation}
Electron-electron interaction is neglected. We solve numerically the
time-independent Schr\"{o}dinger equation
\begin{equation} 
H_S | a \rangle = \epsilon^S_{a} | a \rangle 
\end{equation}
in the basis corresponding to eigenstates of the first three terms in $H_S$
\begin{equation} \begin{split}
&\langle \boldsymbol{r} | nps \rangle = \Psi^S_{nps}(\varphi,z)  
= \sqrt{\frac{1}{r_0 \pi L_0}}\sin{\left(\frac{p\pi z}{L_0}\right)}e^{in\varphi} \chi_s, \\
&\epsilon^S_{nps} = \frac{\hbar^2}{2m_e}\left[ \frac{1}{r_0^2}
\left( n + \frac{\Phi}{\Phi_0} \right)^2 + \frac{p^2 \pi^2}{L_0^2} \right] + \frac{\hbar \omega_c g_e m_e}{4m_0} s.
\label{eq:Basis} 
\end{split} \end{equation}
Here $n \in \mathbb{Z}$ is the orbital angular momentum quantum
number, $s = \pm 1$ describes the spin projection along $z$, 
$\chi_s$ are eigenspinors of $\sigma_z$ and
$p\in \mathbb{Z}_{+}$ arises due to the longitudinal quantization. 

We calculate the charge and current density for a shell conduction 
electron in the state $\langle \boldsymbol{r} |a \rangle = \Psi^S_{a}({\boldsymbol{r}})$ as\cite{Sheng2006}
\begin{equation} 
\begin{split}
&\rho_a ({\boldsymbol{r}}) 
= -e{\Psi^S_{a}}^{\dagger}(\boldsymbol{r})\Psi^S_{a}({\boldsymbol{r}}),\\
&\boldsymbol{j}_a ({\boldsymbol{r}}) 
= \int {\Psi^S_{a}}^{\dagger}(\boldsymbol{r}') \hat{\boldsymbol{j}}(\boldsymbol{r}') \Psi^S_{a}(\boldsymbol{r}') \mathrm{d}\boldsymbol{r}',
\end{split} 
\label{eq:Dens} 
\end{equation}
where the integral extends over the cylinder surface. At vanishing temperature $T\rightarrow 0$ K, placing $N$ electrons
in the system will fill up the $N$ energetically lowest states and the total densities $\rho ({\boldsymbol{r}})$ and $\boldsymbol{j} ({\boldsymbol{r}})$
are obtained by summing up their
contributions. The current density operator is 
\begin{equation} 
\hat{\boldsymbol{j}} (\boldsymbol{r}') = -\frac{e}{2} \left[ \delta (\boldsymbol{r}' - \boldsymbol{r})\boldsymbol{v} + \text{H.c.} \right], 
\label{eq:DensOp} 
\end{equation}
where H.c. denotes the Hermitian conjugate, $\boldsymbol{r}'$ is the coordinate at which the density is
evaluated and $\boldsymbol{r}$ the electron coordinate. The velocity
operator $\boldsymbol{v}$ is determined by solving the Heisenberg equation
of motion, using $H_S$ from Eq.\ (\ref{eq:HS}):
\begin{equation}
\boldsymbol{v} = \left[\frac{p_{\varphi}}{m_e} + \frac{r_0 \omega_c}{2} - \frac{\alpha}{\hbar}\sigma_z \right] \boldsymbol{\hat{\varphi}} + \left[\frac{p_z}{m_e} + \frac{\alpha}{\hbar} \sigma_{\varphi} \right]\boldsymbol{\hat{z}}. \label{eq:vel}
\end{equation}

\subsection{Transport formalism} \label{section:Transport}

In order to calculate the conductance of the finite cylindrical system
we couple it to leads. The leads are taken as cylindrical
continuations of the finite central system with the same radius $r_0$,
extending from the junctions to the left (L) and right (R) along the $z$-axis
over $z<0$ and $L_0<z$, respectively. Their purpose
is to supply phase-coherent electrons to the now open central system
(S) from two reservoirs (or contacts) maintained at chemical potentials
$\mu_L$ and $\mu_R$.\cite{Datta} Electrons propagate through the leads
to the junctions where, as with the central system, hard-wall boundary
conditions are imposed, but injections into the central system are made
possible through a geometry-dependent coupling kernel in the form of an
overlap integral between each lead and the central cylinder.\cite{Gudmundsson2009}
Aside from backscattering due to hard-wall boundary conditions, we assume
that all scattering takes place in the central system. Since $H_S$ in
Eq.\ (\ref{eq:HS}) is time-independent, only elastic scattering is considered.

We define the Hamiltonians and eigenstates 
$H_i | q_i \rangle = \epsilon_{q_i}^i | q_i \rangle$ of the isolated left ($i = L$)
and right ($i = R$) lead, respectively. 
The left and right leads are coupled to the central system by 
assuming coupling terms $H_{LS}$ and $H_{SR}$, respectively, at each junction. 
The two leads are mutually coupled only indirectly through the central system. 
The time-independent Schr\"{o}dinger equation of the entire coupled system 
is written in the matrix form\cite{Thygesen2003,Kurth2005,Brandbygge2002}
\begin{equation}
\begin{bmatrix}
  H_L & H_{LS} & 0 \\
  H_{LS}^{\dagger} & H_S & H_{SR} \\
  0 & H_{SR}^{\dagger} & H_R
 \end{bmatrix}
 \begin{bmatrix}
  | \psi_L \rangle \\
  | \psi_S \rangle \\
  | \psi_R \rangle
 \end{bmatrix}
 =
 E
\begin{bmatrix}
  | \psi_L \rangle \\
  | \psi_S \rangle \\
  | \psi_R \rangle
 \end{bmatrix},
 \label{eq:Sch1}
\end{equation}
where $| \psi_i \rangle$ is the projected ket onto region $i = L,S,R$.
By solving for $| \psi_L \rangle$ and $| \psi_R \rangle$ in the first and third
equations, the retarded Green's operator $G_S$ of the central system
follows from the second equation\cite{Tada2004,Paulsson2007,Paulsson2008}
\begin{equation} 
G_S(E) = \frac{1}{E-H_S - \Sigma_L - \Sigma_R},
\label{eq:GS} 
\end{equation} 
where leads enter through the energy-dependent self-energy operators
\begin{equation} 
\begin{split}
&\Sigma_L = H_{LS}^{\dagger}g_L H_{LS},\\
&\Sigma_R = H_{SR} g_R H_{SR}^{\dagger},
\end{split} 
\label{eq:SelfE}
\end{equation}
for the left and right leads. Note that $G_S$ acts on the central system subspace.
Here, $g_L$ and $g_R$ are the retarded Green's operators of the isolated leads
\begin{equation} 
g_i(E) = \frac{1}{E-H_i + i\eta},
\label{eq:gLeads} 
\end{equation}
where $i = L,R$ and $\eta \to 0^+$. The self-energy operators are generally not
hermitian. Their appearance in $G_S$ motivates the consideration of an
effective central system Hamiltonian $H_S' = H_S + \Sigma_L + \Sigma_R$
which is clearly not hermitian and thus has a complex spectrum.
Provided the self-energies can be regarded as \lq\lq small\rq\rq\
terms compared to $H_S$, the spectrum of $H_S'$ will correspond to the
slightly-shifted spectrum of $H_S$ with added imaginary parts which
provides level-broadening in the central system.\cite{Datta,Kurth2005}

The current through the central system from right lead to left lead is given 
by the (spin-resolved) Landauer formula\cite{Datta,Thygesen2003,Brandbygge2002,Paulsson2008}
\begin{equation} 
I = \frac{e}{h} \int \left( f_L - f_R \right) \text{Tr}\left[ G_S^{\dagger} 
\Gamma_R G_S \Gamma_L \right] \mathrm{d}E,
\end{equation}
where $f_i(E,\mu_i)$ is the Fermi-Dirac distribution function of lead $i$.
The operators $\Gamma_j$ are defined as ($j=L$,$R$)
\begin{equation} 
\Gamma_j = i\left( \Sigma_j - \Sigma_j^{\dagger} \right).
\label{eq:Gamma}
\end{equation}
Assuming that $\mu_L = \mu + \delta \mu$ and $\mu_R = \mu$ where the bias
$\Delta V = \delta \mu/e \rightarrow 0$, at low temperatures
the current becomes linear in $\Delta V$ yielding
the conductance $G$ of the coupled central system in the linear-response
regime\cite{Datta,Thygesen2003,Tada2004}
\begin{equation} 
G(\mu) = \frac{e^2}{h} \text{Tr} \left.\left[ G_S^{\dagger} 
\Gamma_R G_S \Gamma_L \right] \right|_{E=\mu},
\label{eq:G} 
\end{equation}
which will be used to calculate conductance in this paper.
Energy-dependent quantities are evaluated at the chemical potential $\mu$
which is uniform throughout the system. From Eq.\ (\ref{eq:GS}) it is clear
that $G_S$ and hence $G$ are primarily determined by the geometry and
properties of the central system through $H_S$. The leads enter through
the self-energies and provide level-broadening for the central system
as discussed before.

In order to evaluate the trace in Eq.\ (\ref{eq:G}), we construct the
necessary operators in the basis of central system eigenstates $\left\{
| a \rangle \right\}$ in which $H_S$ is diagonal. Specifically,
we construct the matrix representation of $G_S^{-1}$ at a given energy
$E$ with matrix elements 
\begin{equation} \langle a | G_S^{-1} |
b \rangle = \left(E-\epsilon^S_{a}\right)\delta_{a b}
- \langle a | \Sigma_L | b \rangle - \langle a | \Sigma_R
| b \rangle 
\end{equation} 
and invert it numerically. Evaluating the self-energy matrix elements 
$\langle a | \Sigma_i | b \rangle$ with $i = L,R$ also
yields the matrix representations of $\Gamma_i$ in 
Eq.\ (\ref{eq:Gamma}) and is thus sufficient
to calculate $G$ using Eq.\ (\ref{eq:G}).

The matrix elements $\langle a | \Sigma_i | b \rangle$ are
calculated by inserting closure relations for lead $i$ and defining
a coupling kernel between the lead in question and the central system. To illustrate this
procedure, let us consider the left lead matrix element $\langle a |
\Sigma_L | b \rangle$. The normalized eigenstates of the isolated
left lead $\left\{ | q_L \rangle \right\}$ constitute an orthonormal
basis for the state space of the isolated left lead and thus satisfy a
closure relation there. Here, $q_L$ is the set of all necessary orbital and spin
quantum numbers. In this basis, $g_L$ given in Eq.\ (\ref{eq:gLeads}) is
diagonal and $\langle a | \Sigma_L | b \rangle$ can be written as
\begin{equation} 
\langle a | \Sigma_L | b \rangle = \sum\limits_{q_L} \frac{ \langle a |  H_{LS}^{\dagger}
| q_L \rangle \langle q_L | H_{LS}  | b \rangle}{E-\epsilon_{q_L}^L + i\eta}. 
\label{eq:SigLCalc} 
\end{equation}
The coupling Hamiltonian $H_{LS}$ thus enters as an overlap matrix
element between the left lead and the central system eigenstates. Using a coordinate closure
relation for the entire coupled system the overlap matrix element becomes
\begin{equation} 
\langle a |  H_{LS}^{\dagger} | q_L \rangle =
\int\limits_{L,S,R} \mathrm{d}\boldsymbol{r} \mathrm{d}\boldsymbol{r}'
\langle a |  \boldsymbol{r} \rangle \langle \boldsymbol{r} |
H_{LS}^{\dagger} | \boldsymbol{r}' \rangle \langle \boldsymbol{r}' |
q_L  \rangle,
\end{equation} 
where the integrals extend over the central system ($S$) and both leads
($L$,$R$). The functions $\langle \boldsymbol{r} | a \rangle  =
\Psi^S_{a} (\boldsymbol{r})$ and $\langle \boldsymbol{r}' | q_L
\rangle  = \Psi^L_{q_L} (\boldsymbol{r}')$ are localized in the central
system and left lead respectively and vanish everywhere else, so the two
space integrals reduce to integrals over the left lead ($\boldsymbol{r}'$)
and central system ($\boldsymbol{r}$). They couple through
\begin{equation} 
\langle \boldsymbol{r} |  H_{LS}^{\dagger} | \boldsymbol{r}' \rangle 
\equiv K_L(\boldsymbol{r},\boldsymbol{r}'),
\end{equation}
which we define as the coupling kernel between the left lead and the
central system. The overlap matrix element thus becomes
\begin{equation} 
\langle a |  H_{LS}^{\dagger} | q_L \rangle = 
\int\limits_{S} \mathrm{d}\boldsymbol{r} \int\limits_{L} \mathrm{d}\boldsymbol{r}' \left(\Psi^S_{a} 
(\boldsymbol{r})\right)^{\dagger} K_L(\boldsymbol{r},\boldsymbol{r}') \Psi^L_{q_L} 
(\boldsymbol{r}'),
\label{eq:Kern} 
\end{equation}
which can be evaluated with a suitable choice of $K_L(\boldsymbol{r},\boldsymbol{r}')$ 
provided the eigenstates $| a \rangle$ and $| q_L \rangle$ are known.
For lead $i$ we use
\begin{equation} 
K_i(\boldsymbol{r},\boldsymbol{r}') = 
g_0^i e^{- d_z^i |z - z'|} \frac{\delta(\varphi - \varphi')}{r_0},
\label{eq:SpecKern} 
\end{equation}
where $(z,\varphi)$ and $(z',\varphi')$ are coordinates of the central
cylinder and lead $i$, respectively. The kernel is real and due to the
$\delta$-function conserves the angular coordinate between lead and
central system producing a circularly symmetric coupling.
$g_0^i$ is a parameter with the dimension energy/length which
governs the overall strength of the coupling and can be used to control
level-broadening in the central system. The parameter $d_z^i$
determines how rapidly the coupling decreases along the cylinder axis.
To be consistent with the assumption of only indirect coupling between
leads via the central system, $d_z^i$ is chosen such that the exponential
coupling of a given lead vanishes in the vicinity of the other lead. The
$K_i$-modulated overlap integral in Eq.\ (\ref{eq:Kern}) incorporates the
geometry and properties of both central system and leads giving
state-dependent level-broadening.

The eigenstates of the leads are used to calculate $\langle a | \Sigma_i
| b \rangle$. We include the magnetic flux $\Phi$ in the leads and
let it couple to electron spin through the Zeeman term. Hence,
the Hamiltonians $H_L$ and $H_R$ of the left and right leads both have the
form of Eqs.\ (\ref{eq:Kin}) plus (\ref{eq:Zeeman})
with imposed hard-wall boundary conditions at the junctions. 
The eigenstates $H_i | q_i \rangle = \epsilon_{q_i}^i | q_i \rangle$
of the isolated leads are thus characterized by three quantum numbers 
$| q_i \rangle = | n_i k_i s_i \rangle$, $i=L,R$. 
For the left lead
\begin{equation} 
\begin{split}
\langle \boldsymbol{r} | q_L \rangle &= \Psi^L_{n_L s_L k_L} (\boldsymbol{r}) 
= \frac{1}{\pi \sqrt{r_0}} \sin{(k_L z)} e^{in_L \varphi} \chi_{s_L}, \\
\epsilon_{n_Ls_Lk_L}^L &= \frac{\hbar^2}{2m_e r_0^2} \left[ \left(r_0 k_L \right)^2 + 
\left( n_L + \frac{\Phi}{\Phi_0} \right)^2 \right] \\
&+ \frac{\hbar \omega_c g_e m_e}{4m_0}s_L,
\label{eq:EigLeads}
\end{split}
\end{equation}
where $n_L \in \mathbb{Z}$, $k_L \in \mathbb{R}_+$, $s_L=\pm 1$
and $\chi_{s_L}$ is an eigenspinor of $\sigma_z$. Right lead states
$\Psi^R_{n_R s_R k_R} (\boldsymbol{r})$ are obtained by switching the
index $L \rightarrow R$ and setting $z \rightarrow z - L_0$ such that
they vanish at $z = L_0$. 

Using the eigenstates of $H_S$ [Eq.\ (\ref{eq:HS})] and the isolated leads
[Eq.\ (\ref{eq:EigLeads})] along with the kernel of Eq.\ (\ref{eq:Kern}),
the self-energy matrix elements [Eq.\ (\ref{eq:SigLCalc})] are
evaluated. Note that each self-energy matrix element contains two overlap
integrals. The sum over the lead quantum number $n_L$
is truncated at the same value as the central system angular modes $n$
[Eq.\ ($7$)]. For each $n_L$, both spin projections are included and
the integral over the continuous lead quantum number $k_L$ is done
analytically by extension into the complex plane. From the
self-energy matrix elements of both leads at $E = \mu$, the matrix
representations of the operators $G_S(\mu)$ [Eq.\ (\ref{eq:GS})] and
$\Gamma_i(\mu)$ [Eq.\ (\ref{eq:Gamma})] follow and then $G(\mu)$
is calculated using Eq.\ (\ref{eq:G}).

To conclude this section, we mention that alternative, grid-based transport methods exist to calculate
the conductance of nanowires.\cite{Datta} An example is the scattering matrix formalism,
implemented for tubular nanowires in Ref.\ \onlinecite{Serra2009}.

\section{Transport calculations} \label{section:ResultsNoImp}

\subsection{Model parameters}

We consider a cylindrical shell using material parameters for
InAs.\cite{Bringer2011,Snake} The effective electron g-factor is
$g_e = -14.9$ and we use the Rashba SOI parameter $\alpha = 20$
meVnm which corresponds to a strong confining radial field. As the
effective mass of conduction electrons at the $\Gamma$-point we use
$m_e = 0.023m_0$. The dielectric constant is taken as $\epsilon_r =
14.6$. Unless otherwise specified, we assume a shell radius $r_0 = 16.8$
nm and nanowire length $L_0 = 50.4$ nm corresponding to an aspect ratio
$\eta = L_0/r_0 = 3$ which ensures that angular and axial quantization
result in approximately equal level spacing.
Growth of nanowires of comparable radius has been reported in
Refs.\ \onlinecite{Jung2008,Richter2008,Bakkers2003}.

In the following subsections we discuss FP oscillations and
spin effects on the magnetoconductance in our model. We furthermore consider
cylindrical symmetry breaking due to Coulomb impurities and a broken circular
symmetry of the coupling scheme.

\subsection{Flux-periodic oscillations}

\begin{figure}[htbq] 
      \includegraphics[width=0.48\textwidth,angle=0]{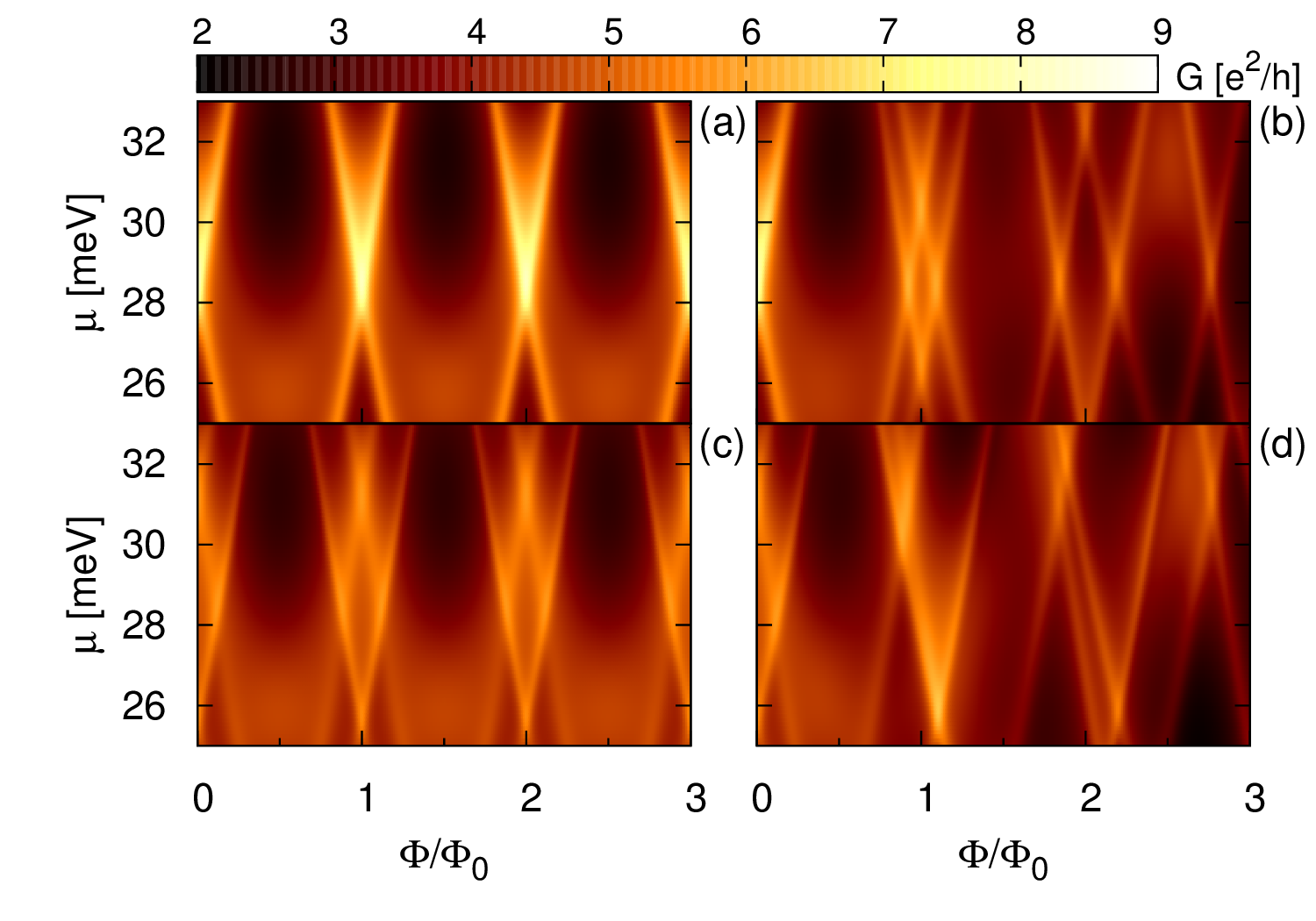}
	\caption{(Color online) Conductance of a cylinder with aspect
	ratio $\eta = 3$ for varying values of $\mu$ and $\Phi$ 
	with: (a) $\alpha = g_e = 0$. (b) $\alpha = 0$, $g_e = -14.9$.
	(c) $\alpha = 20$ meVnm, $g_e = 0$. (d) $\alpha = 20$ meVnm, $g_e
	= -14.9$.  Conductance peaks correspond to broadened chemical
	potential intersections with the spectrum resulting in periodic
	conductance oscillations provided $g_e = 0$ (compare with Fig.\
	\ref{fig:MultiEvB}). Their shape and phase depends on the value
	of $\mu$ considered.}
	\label{fig:ColorCondNoImp}
      \includegraphics[width=0.48\textwidth,angle=0]{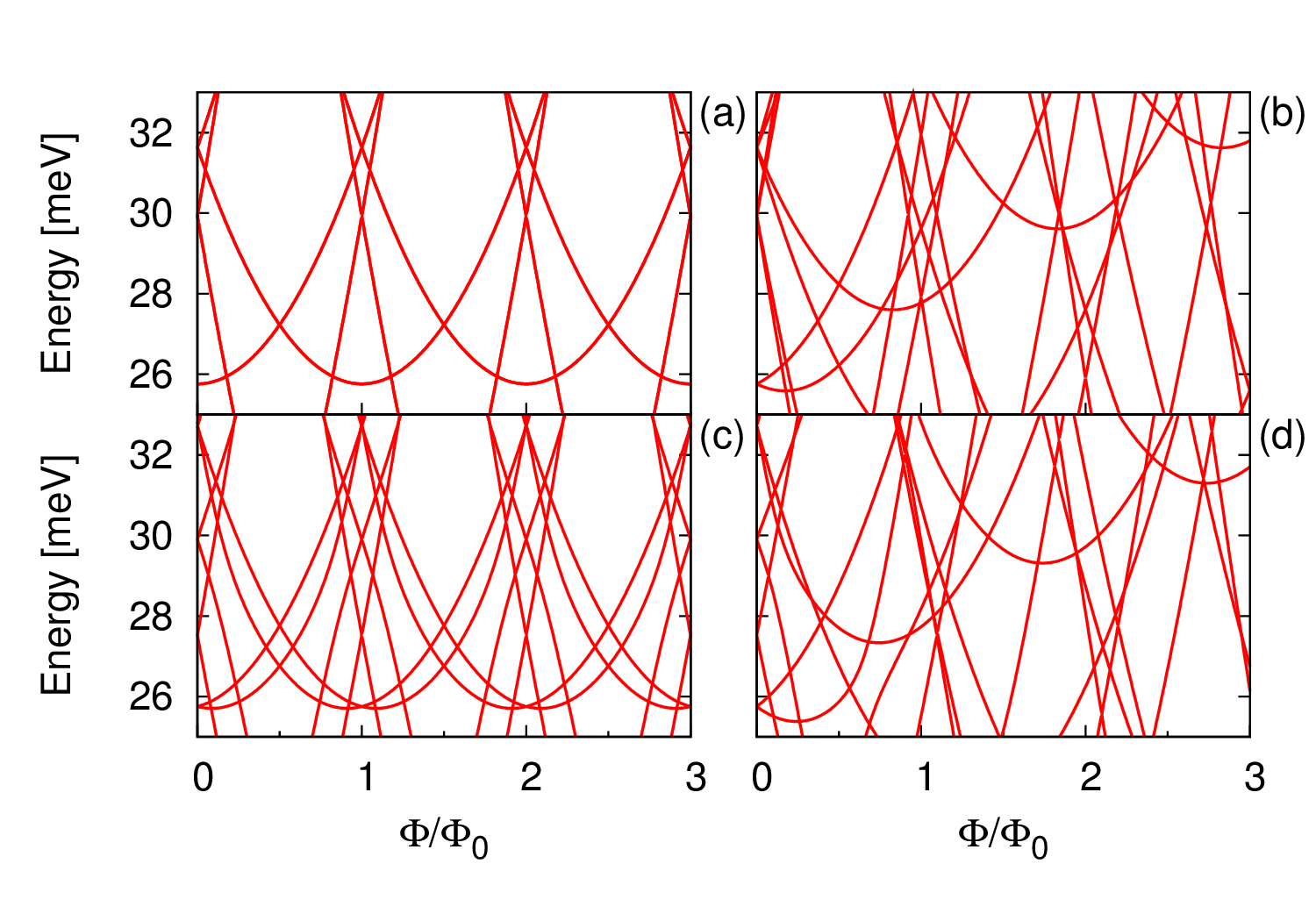}
	\caption{(Color online) Spectrum of a cylinder with aspect ratio
	$\eta = 3$ as a function of $\Phi$ with: (a) $\alpha =
	g_e = 0$. (b) $\alpha = 0$, $g_e = -14.9$. (c) $\alpha = 20$ meVnm,
	$g_e = 0$. (d) $\alpha = 20$ meVnm, $g_e = -14.9$.
	Provided $g_e=0$, the spectrum exhibits periodic oscillations even in the presence of Rashba SOI.}
	\label{fig:MultiEvB}
\end{figure}
Figure \ref{fig:ColorCondNoImp} shows the calculated conductance of a
finite cylinder as a function of $\Phi$ and $\mu$.
Different subfigures demonstrate the effects of the spin-dependent terms
Eqs.\ (\ref{eq:Zeeman}) and (\ref{eq:Rashba}) on $G$.
For reference, the flux-dependence of the closed cylinder spectrum
is given over the corresponding energy range in Fig.\ \ref{fig:MultiEvB}.
A detailed analysis of the spectra of closed cylinders of different aspect ratios is given 
in Ref. \onlinecite{Gladilin2013}. 

Roughly, conductance peaks correspond to $\mu$ intersections with
the spectrum, so $G(\Phi)$ manifests as the broadened and slightly
shifted spectrum $\epsilon^S_{a}(\Phi)$ of $H_S$.
This is due to the self-energy operators of the leads in $G_S$ [Eq.\
(\ref{eq:GS})]. We have deliberately chosen coupling parameters such that
the induced shift and level-broadening are both of the order $1$ meV,
such that the close correspondence between spectrum and conductance in 
Figs.\ \ref{fig:MultiEvB} and \ref{fig:ColorCondNoImp} becomes
evident. Our intention is thus to minimize the effects of the leads and
the particular form of the coupling kernel [Eq.\ ($23$)] on $G$,
which should be governed by the physics of the central system, i.\
e.\ by $H_S$.

In the absence of Rashba SOI, the central system Hamiltonian has the
eigenstates Eq.\ (\ref{eq:Basis}).
At $\Phi = 0$ each level is quadruply degenerate, except
states with $n = 0$ which are only doubly degenerate. At $\Phi = 0$ states
with successively higher orbital angular momentum $L_z = \pm \hbar
n$ pile into a given axial mode forming a ring-like
spectrum until a new axial mode sets in. Hence, the spectrum can be thought of
as a superposition of the ring-like spectra of different axial modes.
When $g_e = 0$ the spectrum is periodic in $\Phi$
with period $\Phi/\Phi_0 = 1$, i.\ e.\ increasing $\Phi/\Phi_0$ by $1$
is equivalent to reducing $L_z$ by $\hbar$ at a fixed energy.\cite{Lorke2000} Hence, the oscillations
are similar to Aharonov-Bohm oscillations which
have been studied extensively in ring-like
\cite{Sheng2006,Takai1993,Takai1994,Washburn1987,Nowak2009,Alexeev2012,Oudenaarden1998,Daday2011,Gefen1984,Buttiker1984}
and cylinder-like
\cite{Holloway2013,Jung2008,Gladilin2013,Tserkovnyak2006,Ferrari2009}
geometries, i.\ e.\ without and with a longitudinal degree of freedom,
respectively. There is no coupling between $\Phi$ and longitudinal
electron motion, so FP oscillations on cylinders manifest due to the
same principles as those observed in quantum rings, but with an added
degree of freedom through the length-dependent $p^2/L_0^2$-term.

\subsection{Zeeman spin effects}

Including the Zeeman term adds the $\Phi$-linear
term $\pm \hbar \omega_c g_e m_e/4m_0 = \pm g_e
(\hbar^2/2m_0r_0^2)(\Phi/\Phi_0)$ to the spectrum such
that the energy of spin down (up) states increases (decreases) with
increasing $\Phi$. Hence spin-degeneracy is lifted as a comparison
between Figs.\ \ref{fig:ColorCondNoImp} (a) and (b) shows. This effect
is pronounced in InAs due to the large value of $g_e$ and ruins the
periodicity of the spectrum.\cite{Gladilin2013,Tserkovnyak2006}
\begin{figure}[htbq]
      \includegraphics[width=0.40\textwidth,angle=0]{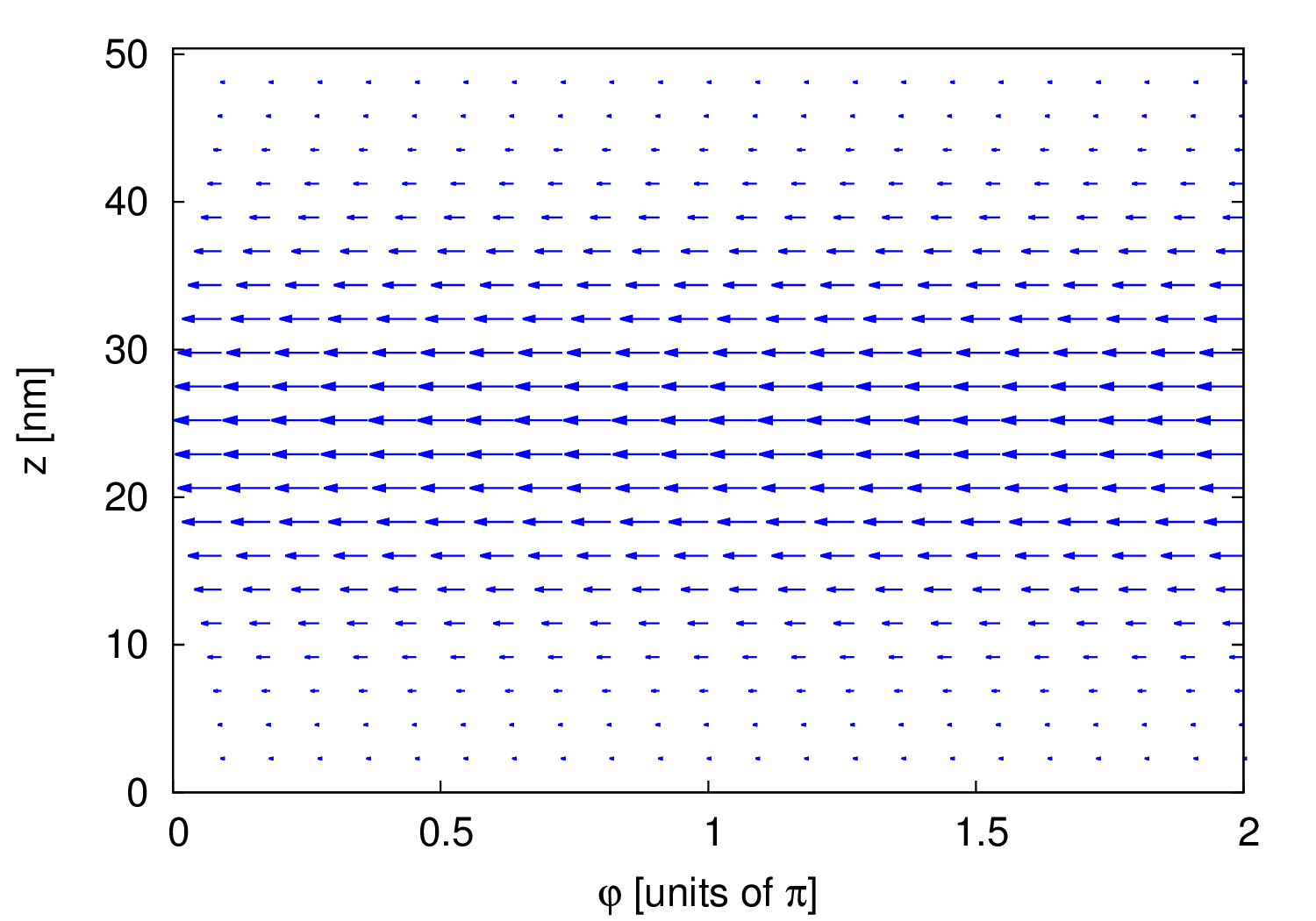}
      \caption{(Color online) Equilibrium current density $\boldsymbol{j}$ on the cylinder surface uncoupled to leads with $\alpha = 0$, pierced by a longitudinal magnetic flux $\Phi/\Phi_0 \approx 0.4$. The current forms concentric circles
      and is circularly symmetric.}
      \label{fig:CDNoRNoZ}
\end{figure}

Figure \ref{fig:CDNoRNoZ} shows the equilibrium current density
$\boldsymbol{j}$ [Eq.\ (\ref{eq:Dens})] on the surface of the
cylinder uncoupled to leads at $\Phi/\Phi_0 \approx 0.4$ with $\alpha = 0$.
The current density is obtained by summing up
the contributions from the $N = 8$ lowest states, a realistic number of
electrons given the density reported in Ref.\ \onlinecite{Bringer2011}.
Setting $g_e = 0$ does not change $\boldsymbol{j}$ in this case, since
the Zeeman term does not affect the velocity operator $\boldsymbol{v}$
[Eq.\ (\ref{eq:vel})]. However, spin-splitting increases with $\Phi$, changing
the orbital characteristics of the energetically lowest states
as they become spin-polarized, which affects $\boldsymbol{j}$.
The system is invariant under rotations around the $z$-axis since
$\left[H_S,D_z(\theta,\boldsymbol{\hat{z}}) \right] = 0$ where $D_z(\theta,\boldsymbol{\hat{z}})
= \exp{\left(i\theta J_z/\hbar \right)}$ is the rotation operator around
the cylinder axis by the finite angle $\theta$. As a result, the electron
density $\rho$ on the cylinder surface is circularly symmetric. Since the
velocity operator $\boldsymbol{v}$ commutes with $D_z$, $\boldsymbol{j}$
is rotationally invariant and because $\langle a | \delta(\boldsymbol{r} -
\boldsymbol{r}')v_z | a \rangle \propto i\langle p | \delta(z - z')\partial_z | p \rangle$
is purely imaginary for all $p$, the axial component
$j_z$ vanishes [Eqs.\ (\ref{eq:Basis}) and (\ref{eq:DensOp})].
Thus $\boldsymbol{j}$ is composed of concentric circles,
each of constant current density. In the closed system, $\rho$ and $\boldsymbol{j}$
thus reflect that electrons enclose a magnetic flux resulting in the
FP oscillations observed in the spectrum. 

In Figs.\ \ref{fig:ColorCondNoImp} (a) and (b) we show the calculated
conductance of a cylinder coupled to leads as a function of $\mu$ and
$\Phi$ without (a) and with (b) the Zeeman term. The flux-dependence
of the spectrum in the corresponding energy range is given in Figs.\
\ref{fig:MultiEvB} (a) and (b). When $g_e = 0$ the conductance evidently
retains the periodicity of the spectrum and hence exhibits oscillations
with period $\Phi/\Phi_0 = 1$. While the conductance oscillations are
periodic for all values of $\mu$, their phase and shape in a single period
is sensitive to the value of $\mu$ considered. Including the Zeeman
term breaks the periodicity of conductance oscillations as it does
to the spectrum.\cite{Tserkovnyak2006} The resulting spin-splitting
of states can produce magnetoconductance curves which are gradually increasing,
decreasing or relatively stable at low values of $\Phi/\Phi_0$ depending
on the value of $\mu$ considered, as may be seen in Fig.\ \ref{fig:ColorCondNoImp} (b).
This point will be further discussed in Sec.\ \ref{section:ExpComp}.
Our numerical results show the same overall trends in the density
of states (DOS) depending on $\mu$ as the flux is varied.

\subsection{SOI effects} \label{section:Rashba}

Including Rashba SOI, we obtain the eigenstates of $H_S$
given by Eq.\ (\ref{eq:HS}) by numerical
diagonalization in the basis (\ref{eq:Basis}). Examples of the resulting energy
spectrum are shown in Figs.\ \ref{fig:MultiEvB} (c) and (d) for $g_e=0$
and $g_e \neq 0$, respectively. Compared to the spectrum with $\alpha = 0$
and $g_e = 0$ in Fig.\ \ref{fig:MultiEvB} (a) the Rashba term generally
lifts spin-degeneracy at $\Phi \neq 0$, but crossings still appear at
integer values of $\Phi$ due to the fact that $H_S$ commutes with $J_z$.

Interestingly, despite Rashba SOI introducing the flux-linear term
$\sim \sigma_z \Phi/\Phi_0$ into the Hamiltonian, the spectrum remains
periodic in $\Phi$. Unlike the Zeeman term, the Rashba term alone does not
break the periodicity of oscillations.\cite{Tserkovnyak2006} Since
$\Phi$ does not couple to $p_z$ in $H_S$, we can look for an
explanation by considering a quantum ring limit $L_0 \rightarrow 0$. This
results in vanishing longitudinal electron motion so $p_z \to 0$
and the term $\sim p_z\sigma_{\varphi}$ vanishes from the Hamiltonian.
The ring-limit spectrum with $g_e = 0$ is
\begin{equation} 
\epsilon^{r}_{ns} = \frac{\hbar^2}{2m_e r_0^2} \left( n + \frac{\Phi}{\Phi_0} \right)^2 - 
\frac{\alpha}{r_0}\left(n + \frac{\Phi}{\Phi_0} \right)s,
\label{eq:RingSpect}  
\end{equation}
which is indeed periodic in $\Phi$ with period $\Phi_0$.
Since the ring-limit and the finite-cylinder spectra couple
identically to $\Phi$, it follows that the spectrum of a finite
cylinder with Rashba SOI alone is periodic in $\Phi$ in agreement
with our numerically obtained spectrum Fig.\ \ref{fig:MultiEvB} (c). Provided $g_e = 0$, the
flux-dependence of the spectrum is qualitatively similar with and
without Rashba SOI aside from the splitting of degenerate states [Figs.\
\ref{fig:MultiEvB} (a) and (c)]. We emphasize that the ring-limit
spectrum with Rashba SOI [Eq.\ (\ref{eq:RingSpect})] differs
from known results for quantum rings. The reason is that in
cylindrical core-shell geometries the confining electric field is
radial,\cite{Bringer2011,Tserkovnyak2006} whereas in quantum rings it is
typically assumed to be perpendicular to the ring, i.\ e.\ along the $z$
direction.\cite{Frustaglia2004,Sheng2006,Nowak2009,Daday2011,Nagasawa2012}

While the Rashba term Eq.\ (\ref{eq:Rashba}) does not commute with $L_z$
or $\sigma_z$, it does commute with $J_z$ and hence with the rotation
operator $D_z$, which means that the Rashba term does not break the
circular symmetry of the system. As a result, the charge and current densities
at a fixed $z$-coordinate remain uniform around the circumference of the
cylinder when Rashba SOI is included. In fact, when $\alpha \neq 0$ the resulting equilibrium current
density of $N = 8$ electrons is almost indistinguishable from that given
in Fig.\ \ref{fig:CDNoRNoZ} where $\alpha = 0$, despite the
presence of SOI-dependent terms in both $j_z$ and $j_{\varphi}$
[Eq.\ (\ref{eq:vel})]. In particular, $j_z$ still vanishes
which suggests that the densities of the observables $p_z$ and $\sigma_{\phi}$
vanish everywhere. Our numerical results show that this is indeed the case.
This is contrary to what happens
on an infinitely long cylinder, where Rashba SOI alone has been shown
to produce a nonvanishing tangential spin density $\sigma_{\phi}$.\cite{Bringer2011}
To understand the difference, let us consider an infinitely long cylinder
$L_0 \rightarrow \infty$. As is shown in Appendix \ref{section:InfCyl},
the normalized eigenspinors of an infinitely long cylinder with Rashba
SOI but $g_e = 0$ can be written as
\begin{equation} \psi_{n}^{\pm}(k) = e^{in\varphi} e^{ikz}
\begin{pmatrix}
  a_{n}^{\pm}(k) \\
  b_{n}^{\pm}(k) e^{i\varphi}
 \end{pmatrix}, \end{equation}
with degenerate energies $E_{n}^{\pm}(k) = E_{n}^{\pm}(-k)$. The
coefficients can be chosen such that they satisfy $a_{n}^{\pm}(-k)
= -a_{n}^{\pm}(k)$ and $b_{n}^{\pm}(-k) = b_{n}^{\pm}(k)$, where
$a_{n}^{\pm}(k)$ is purely imaginary and $b_{n}^{\pm}(k)$ purely
real. From this, it indeed follows that $\langle \psi_{n}^{\pm}(k)
| \sigma_{\varphi} \delta (\boldsymbol{r} - \boldsymbol{r}')
| \psi_{n}^{\pm}(k) \rangle \neq 0$ in agreement with Ref.\
\onlinecite{Bringer2011}. There is however a fundamental difference between
eigenstates on the finite and infinite cylinders, namely that $\langle
p_z \rangle$ always vanishes on the former, but not on the latter
except if $k = 0$. On the finite cylinder with $\alpha \neq 0$,
$\langle p_z \rangle = 0$ because the spectrum of $H_S$ is nondegenerate
at arbitrary values of $\Phi$ and $[H_S,\Pi] = 0$, where $\Pi$ is the spatial
inversion operator over the cylinder center. Hence, the eigenstates $| a \rangle$
have definite parity relative to the cylinder center\cite{Sakurai} and so
$\langle a | p_z | a \rangle = 0$ always since $p_z$ only couples states of
opposite parity. This difference alone implies nonzero $j_z$ on the infinite
cylinder and invalidates a direct comparison between the infinite and finite
systems. But since $E_{n}^{\pm}(k) = E_{n}^{\pm}(-k)$ one can easily construct
infinite-cylinder eigenstates that satisfy $\langle p_z \rangle = 0$
and are thus physically the \lq\lq closest\rq\rq\ ones to finite-cylinder
eigenstates. The most general form satisfying $\langle p_z \rangle = 0$ is
\begin{equation} 
\chi_n^{\pm}(k) = \frac{1}{\sqrt{2}} \left( \psi_n^{\pm}(k) + e^{i\theta} \psi_n^{\pm}(-k) \right), \end{equation}
where $\theta \in \mathbb{R}$. Analogous to the finite
cylinder, one then obtains vanishing tangential spin density $\langle
\chi_n^{\pm}(k) | \sigma_{\phi} \delta (\boldsymbol{r}-\boldsymbol{r}')
| \chi_n^{\pm}(k) \rangle = 0$ which reconciles the two cases.

Figure \ref{fig:ColorCondNoImp} (c) shows the conductance of the
finite cylinder coupled to leads with Rashba SOI included and $g_e
= 0$. Compared with the case $\alpha = g_e = 0$ shown in Fig.\
\ref{fig:ColorCondNoImp} (a), the Rashba term causes a split and shift
of conductance curves. Generally, this results in the appearance
of more peaks of smaller amplitude within a given period at fixed
$\mu$. The splitting and shift is analogous to that which occurs in
the closed-cylinder spectrum [Figs.\ \ref{fig:MultiEvB} (a) and
(c)], further demonstrating the close correspondence between spectrum and
conductance in this formalism. As with the spectrum, including Rashba SOI
alone is insufficient to break the periodic oscillations in conductance
with $\Phi$ at a fixed $\mu$.\cite{Tserkovnyak2006} Instead, it modifies
the shape and phase of conduction curves within a single period. Including
the Zeeman term also breaks the periodicity of the spectrum as in the
case when $\alpha = 0$, see Fig.\ \ref{fig:MultiEvB} (d). Again, this is
reflected in the conductance as Fig.\ \ref{fig:ColorCondNoImp} (d) shows.

\subsection{Broken circular symmetry of the contacts}
\begin{figure}[htbq]
      \includegraphics[width=0.48\textwidth,angle=0]{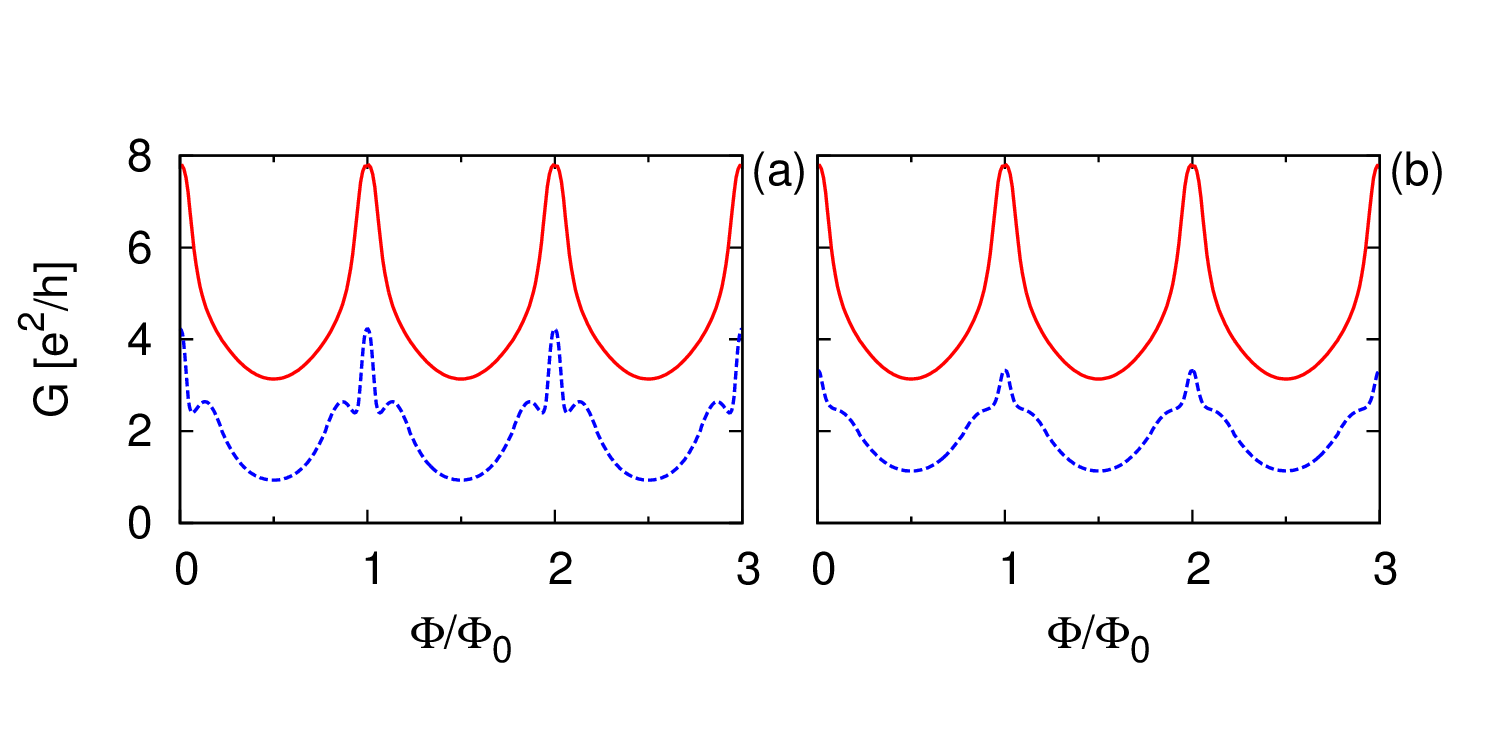}
      \caption{(Color online) Magnetoconductance evaluated at $\mu =
      29$ meV of the cylinder with spin neglected and unrestricted coupling (solid),
      compared to the case with coupling restricted to (dashed): (a) $(\varphi_{min}^L,\varphi_{max}^L)
      = (0,\pi)$ and $(\varphi_{min}^R,\varphi_{max}^R) = (\pi,2\pi)$. (b)
      $(\varphi_{min}^L,\varphi_{max}^L) = (\pi/2,3\pi/2)$ and
      $(\varphi_{min}^R,\varphi_{max}^R) = (\pi/2,2\pi)$. Restricting the coupling
      alters the shape of the conductance oscillations, but they remain flux-periodic.}
      \label{fig:NoCirc}
\end{figure}
From an experimental point of view, the assumption of a circularly symmetric coupling kernel [Eq.\ (\ref{eq:SpecKern})]
may be unrealistic, as contacts typically only connect to restricted parts of the wire.\cite{Gul2014,Blomers2013,Jung2008}
To check whether the FP conductance oscillations are sensitive to this circular symmetry,
we break it explicitly by restricting the coupling regions to finite angles. Assuming vanishing
coupling at junction $i$ except in the angular interval $\varphi_{min}^i \leq \varphi \leq \varphi_{max}^i$, we
introduce step functions into the coupling kernel in Eq.\ (\ref{eq:SpecKern})
\begin{equation} K^i (\boldsymbol{r},\boldsymbol{r}') \rightarrow K^i (\boldsymbol{r},\boldsymbol{r}') \left[\Theta(\varphi - \varphi_{min}^i) - \Theta(\varphi - \varphi_{max}^i) \right]. \end{equation}
Figure \ref{fig:NoCirc} compares the magnetoconductance of the cylinder with restricted and unrestricted coupling. Spin is neglected for simplicity.
We see that the oscillations indeed remain flux-periodic. However, the overall conductance is reduced and the shape of the oscillations within
a given period may change significantly depending on the intervals considered.

\section{Effects of impurities in the core} \label{section:Impurities}

In realistic core-shell nanowires the number of shell conduction electrons may be increased by modulation doping the core with donors.\cite{Blomers2013,Gul2014}
This produces ionized Coulomb impurities in the core, i.\ e.\ attractive potentials to shell conduction electrons. In this section we discuss the
effects of such donor-like impurities on both closed and open cylindrical systems.

\subsection{Coulomb impurities}

Static, donor-like impurities in the core introduce a potential
$V_I$ with which shell conduction electrons interact. The potential
$V_I$ is a sum of individual electron-impurity interaction potentials
$V_I(\boldsymbol{r}) = \sum_{i} \nu_i (\boldsymbol{r})$, where
$\nu_i(\boldsymbol{r})$ is the potential due to impurity $i$ located at
$\boldsymbol{r}_i = (r_i,\varphi_i,z_i)$ given by 
\begin{equation} 
\nu_i (\boldsymbol{r}) = -\frac{e^2}{4\pi\epsilon} \frac{1}{| \boldsymbol{r}
- \boldsymbol{r}_i |},
\label{eq:ImpPot} 
\end{equation} 
where $\epsilon = \epsilon_r \epsilon_0$. The impurities
are accounted for at the single-electron level by adding $V_I$ to
$H_S$ [Eq.\ (\ref{eq:HS})] such that
\begin{equation} H_S = H_O + H_Z + H_R + V_c + V_I. 
\label{eq:HSImp} 
\end{equation}

To obtain the eigenstates with impurities we need to evaluate
the matrix elements of the impurity potential $\langle nps | \nu_i | n'p's' \rangle$ 
in the basis Eq.\ (\ref{eq:Basis}), for which we use a convenient
expansion of the three-dimensional Coulomb potential in cylindrical
geometries.\cite{Cohl1999} It can be written as
\begin{equation} 
\frac{1}{| \boldsymbol{r} - \boldsymbol{r}_i |} = 
\sum\limits_{m = -\infty}^{\infty} e^{im(\varphi - \varphi_i)} 
\frac{1}{\pi\sqrt{r_0 r_i}} Q_{m-1/2}\left( \gamma \right),
\label{eq:CoulExp} 
\end{equation}
where $Q_{m-1/2}$ are associated Legendre functions of the second kind of
zeroth order and half-integer degree and $\gamma = [r_0^2 + r_i^2 +
(z-z_i)^2]/2r_0r_i$. The Legendre functions
are obtained using the code provided in Ref.\ \onlinecite{Segura2000}.

Alternatively one may omit $V_I$ from $H_S$ and instead introduce the impurities directly into the central system Green's function by solving the Dyson equation
\begin{equation} 
G_S^I = G_S + G_SV_IG_S^I. 
\label{eq:Dyson} 
\end{equation}
Solving it yields the Green's function $G_S^I$ of the central system,
where both leads and impurities are accounted for, from the Green's
function $G_S$ given in Eq.\ (\ref{eq:GS}). Impurities are then included
in calculations of $G$ with Eq.\ (\ref{eq:G}) using the same formalism
and procedure as discussed in Sec.\ \ref{section:Transport}, but
replacing $G_S$ with $G_S^I$ everywhere. Comparing these two methods, we find that
they give the same results. 
To conclude this subsection, we remark that
Eq.\ (\ref{eq:Dyson}) is generally not solvable by
iteration, as the matrix $G_S V_I$ can have a spectral radius $\rho_T >
1$ which makes the iteration scheme nonconvergent. Instead, we use $G_S
(G_S)^{-1} = I$ and rewrite the Dyson equation as
\begin{equation} 
\left[ \left( G_S \right)^{-1} - V_I \right]G_S^I = I,
\end{equation}
which we solve numerically as a system of equations.

\subsection{Particle and current densities}

\begin{figure}[htbq]
      \includegraphics[width=0.48\textwidth,angle=0]{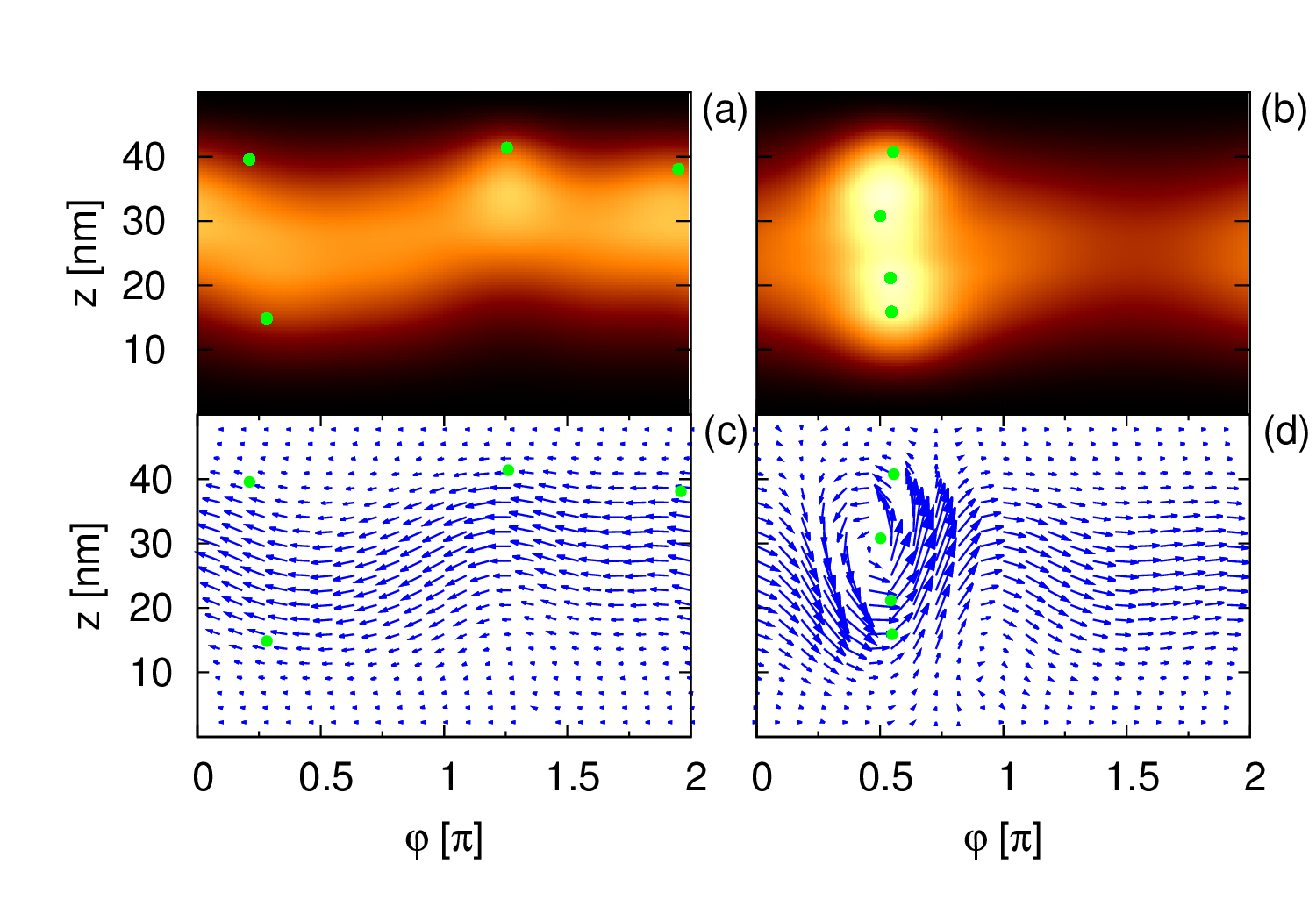}
	\caption{(Color online) Electron (top) and current (bottom) densities of $8$ electrons on the cylinder with aspect ratio $\eta = 3$ pierced by a longitudinal flux $\Phi/\Phi_0 \approx 0.4$ with two, separate impurity configurations (left and right).
	Due to the impurities (filled dots) the rotational and parity symmetries are broken (compare with Fig.\ \ref{fig:CDNoRNoZ}).
	Bright and dark regions correspond to regions of high and low charge density, respectively.}
	\label{fig:DensImp}
\end{figure}

Impurity potentials of the form Eq.\ (\ref{eq:ImpPot}) 
break the circular symmetry of the system, except in the special case when
the impurities lie on the cylinder axis $r_i = 0$.\cite{Gladilin2013}
Small deviations in impurity location from the cylinder axis 
introduce in the spectrum avoided crossings for states with low $L_z$.
The gaps are small if only few impurities are present and located close
to the cylinder axis. Avoided crossings in rings due to disorder are
for example discussed in Ref.\ \onlinecite{Buttiker1984}.
The impurities also couple to longitudinal electron motion and so
their location on the $z$-axis can strongly affect densities, as the
impurity potentials generally ruin also the longitudinal parity symmetry
of the system. For example, if the impurities are concentrated close
to the upper end $z=L_0$ of the cylinder, the longitudinal symmetry is
manifest broken as $\boldsymbol{\rho}$ and $\boldsymbol{j}$ increase
in the upper half but decrease in the lower half. Placing impurities
close to the cylinder center ($r_i = 0$, $z_i = L_0/2$) will however
produce densities that are nearly indistinguishable from the case
without impurities.

Impurities are generally not located solely around the center of the
cylinder axis in realistic core-shell nanowires and densities
may differ significantly for more generalized distributions. In
Fig.\ \ref{fig:DensImp} we show the densities for two distributions,
where impurity coordinates $(\varphi_i,z_i)$ are marked with large
dots. The calculations are done with spin suppressed. The impurities
in Figs.\ (a) and (c) (configuration 1) are uniformly
distributed along the radial direction with coordinates ranging between
$0.15 \leq r_i/r_0 \leq 0.82$, more concentrated in the upper half of
the cylinder. In Figs.\ (b) and (d) (configuration 2) the impurities
are condensed into a narrow angular interval around $\varphi_i \approx
\pi/2$ close to the cylinder surface with $0.55 \leq r_i/r_0 \leq 0.76$. Both
configurations strongly break the rotational and parity symmetries in
the closed system as the densities show. Configuration 2 is composed
of impurities that are evenly distributed along the cylinder length at
comparable distances from the surface in a narrow angular interval. As a
result, they form a potential well around $\varphi \approx \pi/2$ which
traps states of low orbital angular momentum $L_z$. This \lq\lq flattens\rq\rq\
the corresponding energy levels as functions of $\Phi$ and suppresses their FP
oscillations, similar to a transverse electric field.\cite{Barticevic2002,Alexeev2012}
This is reflected in $\boldsymbol{j}$ [Fig.\ \ref{fig:DensImp} (d)] which
shows the formation of a vortex circulating the impurity cluster, greatly
deforming the circular motion. As Fig.\ \ref{fig:DensImp} (c) shows, configuration 1
affects $\boldsymbol{j}$ more modestly by for example
introducing nonvanishing $j_z$.

\subsection{Magnetoconductance with impurities}
\begin{figure}[htbq]
      \includegraphics[width=0.48\textwidth,angle=0]{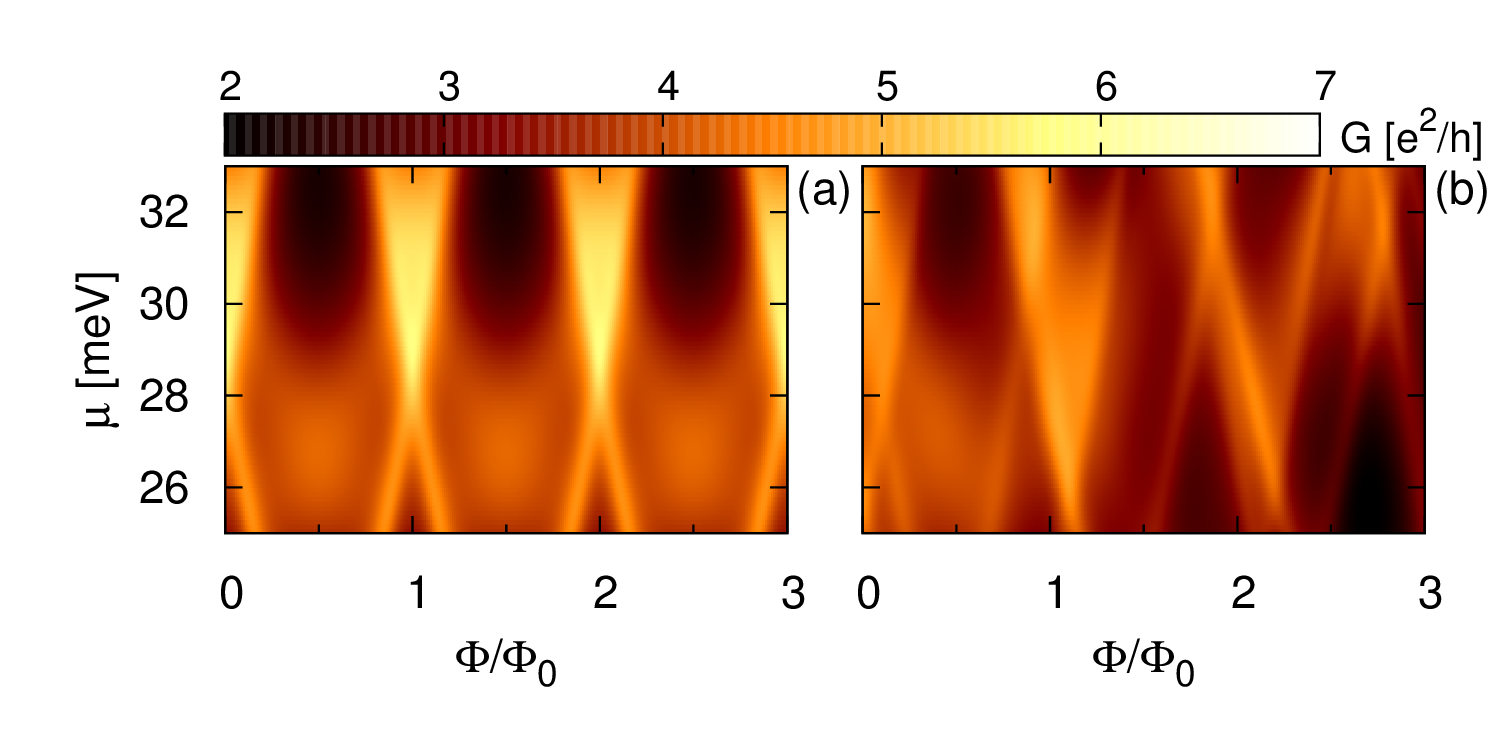}
      \caption{(Color online) Magnetoconductance of a cylinder with aspect ratio $\eta = 3$ with impurity configuration 1
      [Figs.\ \ref{fig:DensImp} (a) and (c)] without (a) and with (b) spin included. The figures are qualitatively similar to the case without
      impurities given in Figs.\ \ref{fig:ColorCondNoImp} (a) and (d), but with damped oscillations. Impurities alone are insufficient to break the periodicity
      of the oscillations.}
      \label{fig:ImpCond}
\end{figure}

Donor-like impurity potentials are attractive to electrons and will thus
shift the spectrum of the central system down in energy in addition to
deforming the $\Phi$-dependence. Both factors depend on the number
and location of impurities and furthermore different states may be
affected differently. As conductance is evaluated at a fixed $\mu$
set by the leads and primarily determined by the spectrum of $H_S$,
adding impurities can significantly alter $G(\Phi)$ at a fixed $\mu$
solely due to the induced shift of the spectrum. We use a model
gate voltage
\begin{equation} 
H_S \rightarrow H_S + eV_G
\label{eq:Gate}
\end{equation}
to shift the central system spectrum for a given configuration so that ground
state energies match with and without impurities, in order to 
make possible a comparison between different impurity configurations at the same
chemical potential. To demonstrate the effects of impurities
on magnetoconductance, let us consider impurity configuration 1 used in
Figs.\ \ref{fig:DensImp} (a) and (c). Realigning the spectrum requires a
gate voltage $V_G = 19.1$ mV. Figure \ref{fig:ImpCond} shows the resulting
conductance of the finite cylinder coupled to leads without (a) and with
(b) spin included as a function of $\Phi/\Phi_0$ and $\mu$. Provided the
oscillations were periodic prior to the inclusion of impurities (i.\ e.\
if $g_e = 0$) they remain so when impurities are included. 
This is because the impurity potentials do
not couple to $\Phi$. The conductance curves with and without impurities
[Figs.\ \ref{fig:ColorCondNoImp} (a) and (d)] are qualitatively similar,
but the impurities dampen oscillation amplitudes by reducing maxima and
increasing minima.
\begin{figure}[htbq]
      \includegraphics[width=0.48\textwidth,angle=0]{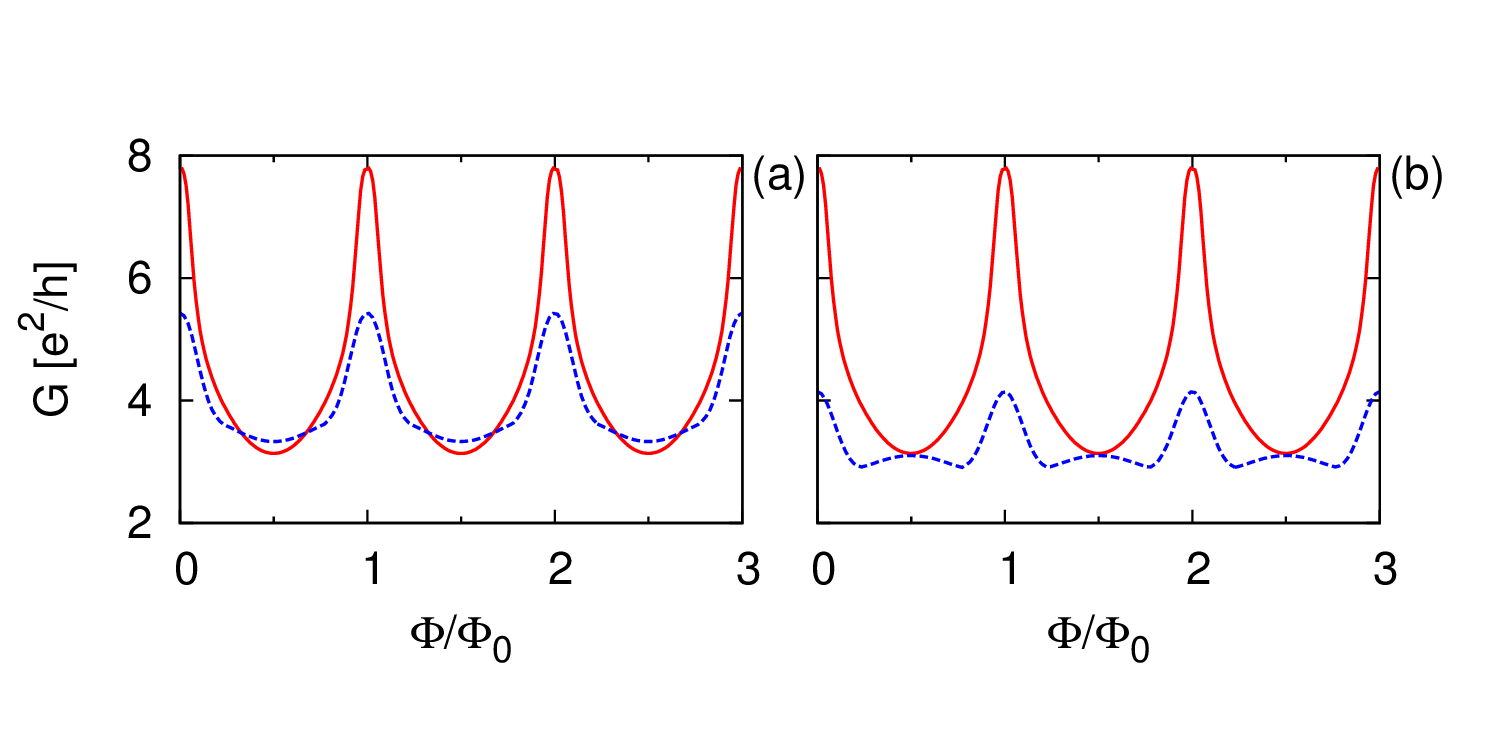}
      \caption{(Color online) Magnetoconductance evaluated at $\mu =
      29$ meV of the cylinder with spin neglected, averaged over $N_c =
      250$ (dashed) random impurity configurations
      containing (a) $N = 4$ and (b) $N = 8$ impurities each.
      Further averaging does not affect the results significantly. The solid
      lines show $G(\Phi)$ without impurities. Impurity averaging reduces
      conductance oscillations, but even with a highly doped core ($N =
      8$) they are still clearly visible.}
      \label{fig:ImpAvg}
\end{figure}

So far, we have considered the conductance of
particular impurity configurations and seen that FP oscillations can survive
in the presence of impurity-induced dampening. We can also evaluate the average
magnetoconductance $\langle G(\Phi) \rangle$ at a fixed $\mu$ over
multiple random impurity configurations, which gives insight into the
general behavior of an assembly of core-shell nanowires. We calculate
$\langle G(\Phi) \rangle$ over $N_c$ random configurations of $N$
impurities each, where $N$ is constant for a given assembly. The
assumption of a constant number of impurities per configuration is
justified using reported average donor densities in the core. For
reference, a core donor density of $10^{17}$ cm$^{-3}$, which is
large assuming a GaAs core, corresponds to $4$ or $5$ impurities in
the central system under consideration.\cite{Gul2014,Blomers2013}

Figure \ref{fig:ImpAvg} (a) compares $G(\Phi)$ at $\mu = 29$ meV
without impurities and $\langle G(\Phi) \rangle$ averaged over $N_c = 250$
configurations of $N = 4$ impurities. We neglect spin for simplicity.
Further averaging does not affect the results significantly. The applied gate voltage $V_G$
is obtained by averaging the shift of the ground state over multiple
$N$-impurity configurations. The oscillations are indeed damped,
but present. Increasing the number of impurities to $N = 8$ [Fig.\
\ref{fig:ImpAvg} (b)] the amplitude drops, but the oscillations
still survive. Actually, for $N = 8$ impurities, even averaging over
$N_c = 10$ configurations already yields qualitatively the same $\langle
G(\Phi) \rangle$ as observed in Fig.\ \ref{fig:ImpAvg} (b) after extensive
averaging, which implies that at such high core donor concentrations the
exact impurity configuration is not paramount. The damping suggests that
conductance oscillations may be reduced in amplitude beyond achievable
experimental resolution in extremely disordered samples. However, our
simulations indicate that even in the presence of a large but realistic
core-donor density, the oscillations are clearly resolvable. Finally,
we mention that our model donor impurities [Eq.\ (\ref{eq:ImpPot})] do
not account for screening. Screening of
donor impurities in the core would reduce their effects on conduction
electrons and hence on both closed and open system properties. Similarly,
electron-electron interaction would oppose impurity-induced localizations
in the system, e.\ g. as in Figs.\ \ref{fig:DensImp} (b) and (d), and hence weaken
impurity effects.\cite{Gladilin2013} By ignoring screening effects
in the core and electron-electron interaction, our simulations thus
describe \lq\lq a worst-case scenario\rq\rq\ of the electron-impurity
interactions.

\section{Comparison with recent experimental data} \label{section:ExpComp}

In this section we compare simulations using realistic parameters with recently reported
measurements performed on GaAs/InAs core-shell nanowires. Results in
Ref.\ \onlinecite{Gul2014} show FP conductance oscillations superimposed
on slowly-varying background oscillations in hexagonal GaAs/InAs
core-shell nanowires. The background oscillations are attributed to
universal conductance fluctuations. Our cylindrical model represents
an idealized core-shell nanowire and neglects some aspects present in experiment,
notably the hexagonal structure, shell thickness and electron-electron interaction.
Out of the effects considered in this paper, we have
shown that only Zeeman splitting can break the periodicity of
oscillations in cylinders, e.\ g.\ Fig.\ \ref{fig:MultiEvB}.
\begin{figure}[htbq]
      \includegraphics[width=0.48\textwidth,angle=0]{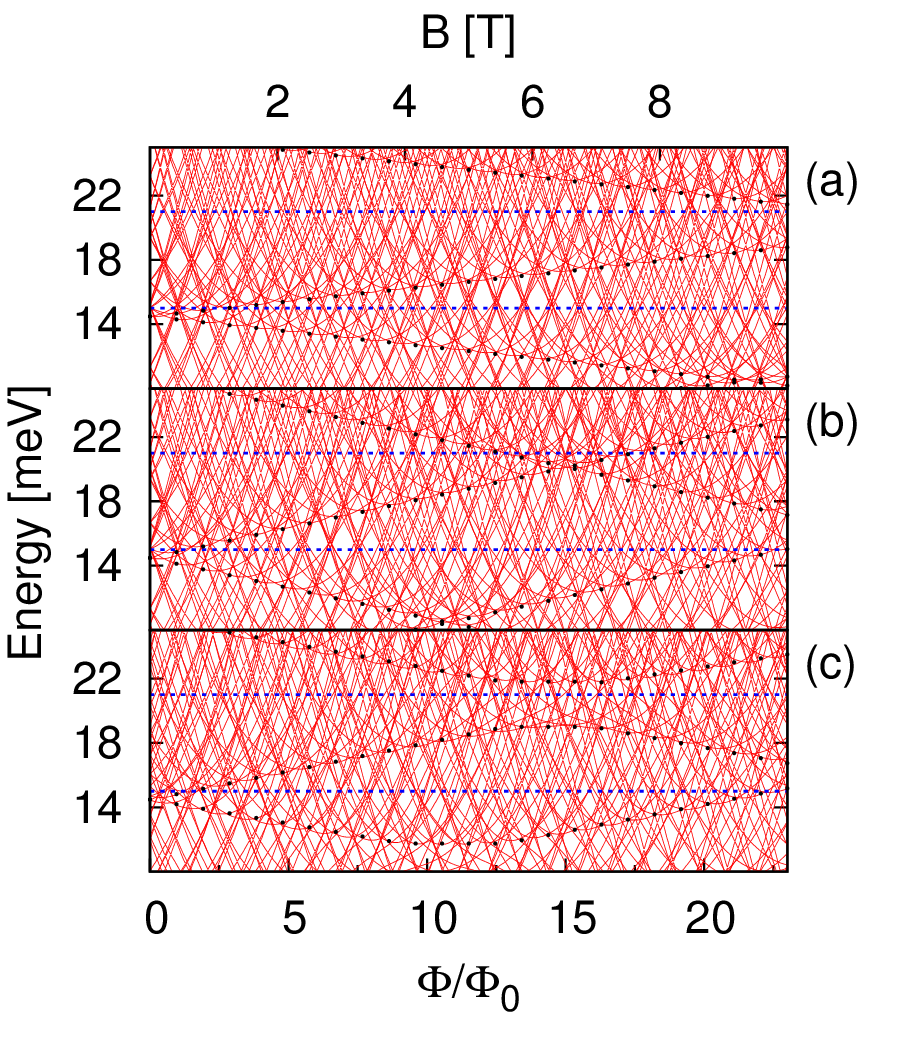}
      \caption{(Color online) Spectrum of a cylinder with $r_0 = 55$ nm and $L_0 = 100$ nm with: (a) $g_e = -14.9$, $\alpha = 0$. (b) $g_e = -29.8$, $\alpha = 0$.
      (c) $g_e = -29.8$, $\alpha = 20$ meVnm. Due to Zeeman splitting, axial-sublevel minima produce sloped linear \lq\lq traces\rq\rq\ of parabola minima,
      marked with dots, resulting in large-scale DOS variations at a fixed energy. Increasing $g_e$ amplifies this effect and reveals crossings between traces.
      With Rashba SOI included the crossings become avoided. The values of $\mu$ used to calculate $G(\Phi)$ in Fig.\ \ref{fig:sim3} are
      marked with horizontal dashed lines.}
      \label{fig:sim2}
\end{figure}
Consider a cylinder with $r_0 = 55$ nm and $L_0 = 100$ nm. Figure
\ref{fig:sim2} (a) shows the energy spectrum with $g_e = -14.9$ and $\alpha
= 0$. Each axial mode has an energy minimum,
the spin-degeneracy of which is lifted by the Zeeman term for $\Phi \neq
0$, producing sloped \lq\lq traces\rq\rq\ of the corresponding parabolic
bottoms marked with filled circles, yielding a flux-modulated DOS at a fixed energy.
The energies $15$ and $21$ meV (dashed) are located between two such
traces approaching the former and distancing from the latter, resulting
in a monotonically increasing and decreasing DOS, respectively. This is
reflected in $G(\Phi)$, which Fig.\ \ref{fig:sim3} (a) shows evaluated
at the corresponding values $\mu = 15$ and $21$ meV, decreasing and
increasing gradually on average, comparable to the experimental results of Ref.\
\onlinecite{Gul2014}. It follows that the measured background oscillations
could be explained as an interplay between the finite system length
and spin. To further illustrate this effect, Fig.\ \ref{fig:sim2} (b)
shows the cylinder spectrum with double Zeeman interaction, $g_e =
-29.8$, and $\alpha = 0$ still. The slopes of the parabola-minima traces
increase revealing crossings and the DOS modulation is amplified.
The DOS is maximum when two such bottom-band traces cross,
i.\ e. for $\Phi/\Phi_0 \approx 15$ just below $21$ meV, and minimum at the largest energy
separation between them.
Figure \ref{fig:sim3} (b) shows the corresponding conductance for $\mu =
15$ and $21$ meV and reveals that the crossings manifest as peaks in background
conductance oscillations.

We can apply the correspondence between crossings of traces
and peaks in background oscillations to calculate the electron g-factor.
A trace is formed as a function of $\Phi$ by the energy minima of a given
axial mode, described by the spectrum Eq.\ (\ref{eq:Basis}) with $n = -\Phi/\Phi_0$
which yields an equation for lines, namely the traces. An intersection between two traces occurs
at a particular value of $\Phi$ when the two corresponding lines intersect.
Solving for $g_e$ yields
\begin{equation}
|g_e| = \frac{\pi^2 m_0 r_0^2}{2m_e L_0^2} \left( \frac{\Phi}{\Phi_0} \right)_c^{-1} |p_2^2 - p_1^2|, \label{eq:geval}
\end{equation}
as $s_1 - s_2 = \pm 2$ because only traces of opposite spin may intersect.
Here, $(\Phi / \Phi_0 )_c$ is the magnetic flux at
which the lines intersect. It may be estimated from Fig.\ \ref{fig:sim3}
as the flux at which the background
conductance oscillations peak. Applying this to $G$ at $\mu = 21$ meV in
Fig.\ \ref{fig:sim3} (b), we find $(\Phi / \Phi_0 )_c \approx 15$.
If the chemical potential is known, the axial modes follow
from the condition $p_1^2 < 2m_eL_0^2 \mu /(\hbar \pi)^2 <p_2^2$, which must hold at $\Phi = 0$.
For the present example, we find $p_2 = 4$ and $p_1 = 3$ which yields $g_e \approx -30$
compared to the input value $g_e = -29.8$.
If the chemical potential is not known, a guess of the relevant axial modes is needed.

In Figs.\ \ref{fig:sim3} (a) and (b)
we also note a beating pattern in the conductance, which in our model arises due to Zeeman
splitting causing a misalignment of the flux-parabolas at a fixed energy [Eq.\ (\ref{eq:Basis})].
Doubling the g-factor results in a smaller beating period as a comparison between the curves at $\mu = 15$ meV clearly illustrates.
Beating patterns are observable in experiment, but we note that they may also be caused by other mechanisms than spin splitting.
For example, electrons might occupy higher radial modes in a shell of finite thickness and hence have different
effective radii, such that more than one magnetic flux is distinguishable. The superposition of the corresponding oscillations,
which are periodic in their respective fluxes, would produce a beating pattern. However, previous calculations indicate
that the oscillations should remain periodic in the presence of a small, nonzero shell thickness.\cite{Gladilin2013,Tserkovnyak2006}

\begin{figure}[htbq]
      \includegraphics[width=0.48\textwidth,angle=0]{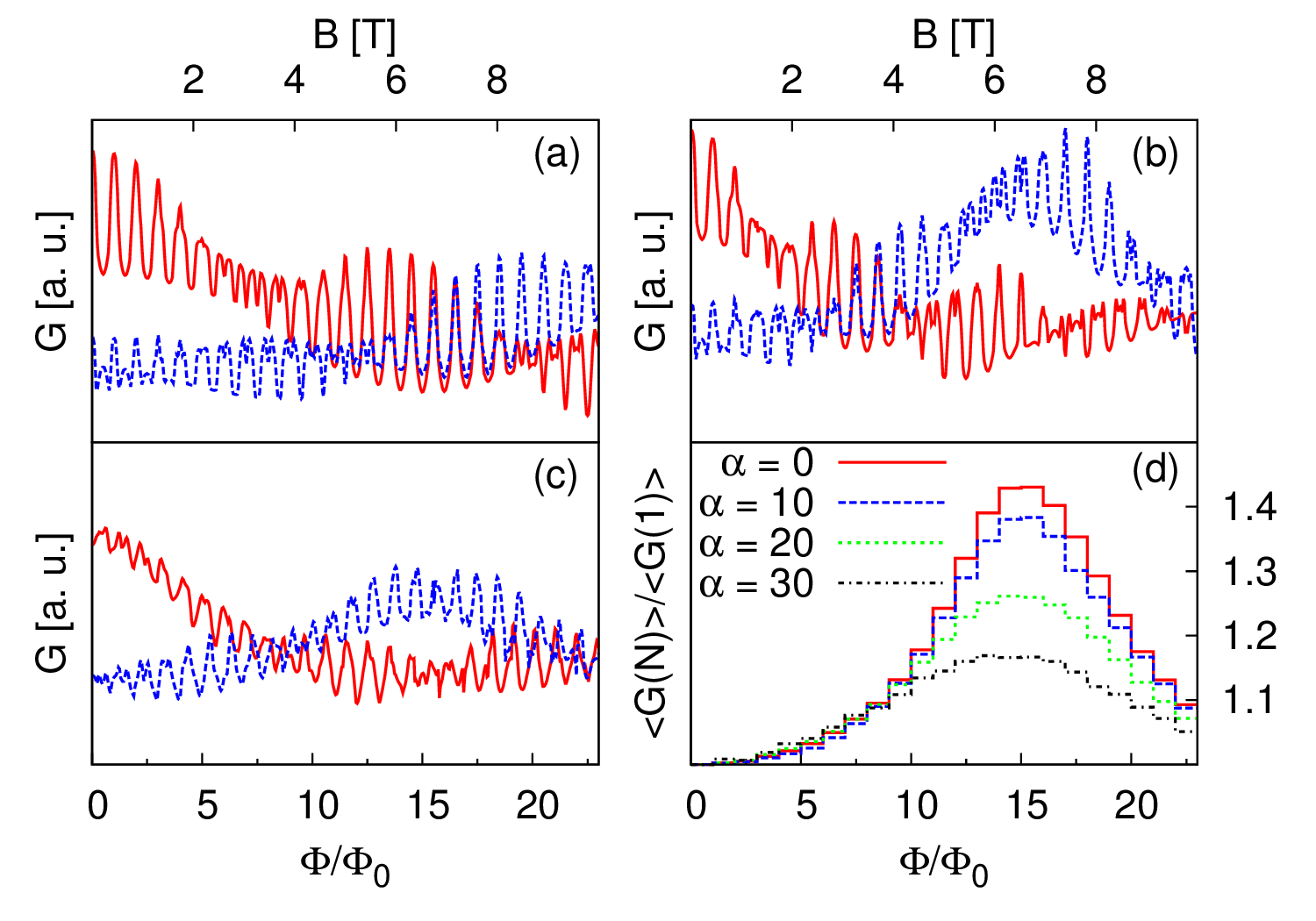}
      \caption{(Color online) A cylinder with $r_0 = 55$ nm and $L_0 = 100$ nm. (a), (b) and (c): $G(\Phi)$ evaluated at $\mu=15$ meV (solid) and $\mu = 21$ meV (dashed)
      with $g_e = -14.9$, $\alpha = 0$ (a), $g_e = -29.8$, $\alpha= 0$ (b) and $g_e = -29.8$, $\alpha = 20$ meVnm (c). Due to Zeeman splitting, conductance oscillations
      are superimposed on background fluctuations, the form of which depends on $\mu$ as is reflected in the spectrum [compare with Figs.\ \ref{fig:sim2} (a), (b) and (c)].
      (d) Flux-averaged conductance $\langle G(N) \rangle$ relative to $\langle G(N=1) \rangle$ at $\mu = 21$ meV plotted against flux number $N$ with $g_e = -29.8$
      for different values of $\alpha$. As $\alpha$ increases, the amplitude of the peak around $\Phi/\Phi_0 = N = 15$ is reduced, reflected in the Rashba-induced
      avoided crossings of \lq\lq traces\rq\rq\ in Fig.\ \ref{fig:sim2} (c).}
      \label{fig:sim3}
\end{figure}
Finally, Fig.\ \ref{fig:sim2} (c) shows the spectrum with
$g_e = -29.8$ and $\alpha=20$ meVnm. Interestingly, due to the SOI
the crossings of the traces become avoided, 
their energy separation increasing with $\alpha$. The resulting
energy \lq\lq gap\rq\rq\ dampens the background-oscillation peaks of $G(\Phi)$,
as is shown for $\mu = 21$ meV at $\Phi/\Phi_0 \approx 15 $ in Fig.\ \ref{fig:sim3} (c). 
Rashba SOI also dampens the FP oscillations themselves as discussed in Sec.\ \ref{section:Rashba}.
To understand how the amplitude of the background conductance oscillations varies with $\alpha$, Fig.\ \ref{fig:sim3}
(d) shows how the conductance $\langle G(N) \rangle$ averaged over the
$N$-th flux $N-1 \leq \Phi/\Phi_0 \leq N$ with $N \in \mathbb{Z}_+$
varies with $N$ relative to $\langle G(N=1)\rangle$ for different values
of $\alpha$. By averaging over the intervals between integer 
fluxes we exclude the
$\Phi$-periodic part of $G$ and isolate the Zeeman-induced 
background oscillations. In analogy with Figs.\ \ref{fig:sim3} (b) and (c),
$\langle G(N) \rangle$ peaks around $N = 15$ and as $\alpha$ increases
the peak is reduced in amplitude relative to $\langle G(1)\rangle$. It
has been shown\cite{Engels1997,Nitta1997,Liang2012} that $\alpha$ is controllable
by applying a gate voltage and therefore measurements on peaks in the
background oscillations of magnetoconductance in GaAs/InAs core-shell
nanowires may allude to the existence of Rashba SOI in such tubular
systems. Importantly, for $\alpha \neq 0$ the background conductance oscillations flatten, but the peaks
do not shift much compared to $\alpha = 0$,
and so Eq.\ (\ref{eq:geval}) may still be applied to estimate $g_e$.

\section{Conclusions} \label{section:Conclusions}
We performed transport calculations of electrons situated on a cylindrical
surface in the presence of a longitudinal magnetic flux and obtained flux-periodic oscillations
at different chemical potentials. Varying $\mu$ shifts the chemical potential of the system
relative to the fixed spectra of the central part and the leads, similar to the experimental setup in
Ref.\ \onlinecite{Gul2014}, where both the nanowire and the contacts are placed on a substrate
used as a back gate. An alternative model is to shift only the central system spectrum relative
to the leads and some fixed chemical potential, but our calculations (not shown here) reveal
that these two methods are essentially identical, with only a minor difference in level-broadening.
The oscillations survive and remain periodic in the presence of
impurities and occur even if the contacts do not have a uniform angular
coverage of the cylindrical surface. Hence, they are robust to deviations
from the ideal circular and parity symmetries in the nanowire. Furthermore, the oscillations remain flux-periodic
when Rashba SOI is included. The oscillations are also still present when
Zeeman interaction is included, although they cease to be flux-periodic. Instead,
a rich structure of beating patterns and background oscillations is identified,
the latter of which also relates to the finite system length. By analyzing these
oscillations, it is possible to estimate the g-factor of the electrons in the
shell and detect the presence of Rashba SOI, provided the SOI strength can be varied.
Our results are in qualitative agreement with recent measurements on GaAs/InAs core-shell
nanowires.

\begin{acknowledgments}
We would like to thank Thomas Sch\"{a}pers, Fabian Haas, Sigurdur Ingi Erlingsson,
Llorens Serra and Thorsten Arnold for enlightening discussions. This work was supported
by the Research Fund of the University of Iceland and the Icelandic Research and Instruments Funds.
\end{acknowledgments}

\appendix
\section{Infinite cylinder eigenstates with Rashba SOI} \label{section:InfCyl}
Consider an infinitely long cylinder with Rashba SOI pierced by a longitudinal magnetic flux but no Zeeman coupling, i.\ e.\ with Hamiltonian $H_S^I = H_O + H_R$.
Since $\left[H_S^I, p_z \right] = \left[H_S^I, J_z \right] = 0$ we look for spinor solutions of the form \cite{Bringer2011,Mehdiyev2009}
\begin{equation} \psi_{nk}(z,\varphi) = \frac{1}{2\pi\sqrt{r_0}}e^{ikz} e^{in\varphi}\binom{a}{be^{i\varphi}} \end{equation}
where $n\in\mathbb{Z}$ and $k \in \mathbb{R}$. The time-independent Schr\"{o}dinger equation $H \psi_{nk} = E \psi_{nk}$ yields
\begin{equation}
\begin{bmatrix}
  \xi (n+\tilde{\Phi})^2 - \tilde{E} - \frac{\alpha}{r_0}(n + \tilde{\Phi} ) \hspace{0.2 cm} ; \hspace{0.2 cm} -i\alpha k \\
  i\alpha k \hspace{0.2 cm} ; \hspace{0.2 cm} \xi (n+1+\tilde{\Phi})^2 - \tilde{E} + \frac{\alpha}{r_0}(n + 1 + \tilde{\Phi} )
\end{bmatrix}
\begin{pmatrix}
  a \\
  b 
 \end{pmatrix}
 = 0
\end{equation}
where we define $\xi = \hbar^2/2m_e r_0^2$, $\tilde{\Phi} = \Phi/\Phi_0$ and $\tilde{E} = E - \xi k^2 r_0^2$.
The energies are
\begin{equation} E_{n}^{\pm}(k) = \xi k^2r_0^2 + \frac{1}{2}\left(A \pm \sqrt{A^2 - 4(B - \alpha^2 k^2)} \right)  \end{equation}
where $A(n,\tilde{\Phi})$ and $B(n,\tilde{\Phi})$ are independent of $k$ such that $E_{n}^{\pm}(k) = E_{n}^{\pm}(-k)$.
To normalize, the spinor coefficients $a$ and $b$ can be chosen as
\begin{equation} a^{\pm}_n(k) = \frac{i\alpha k}{\xi (n+\tilde{\Phi})^2 - \frac{\alpha}{r_0}(n+\tilde{\Phi}) - \tilde{E}^{\pm}_n(k)}b^{\pm}_n(k) \end{equation}
and
\begin{equation} b^{\pm}_n(k) = \sqrt{\left( 1 + \frac{\alpha^2 k^2}{\left[\xi (n+\tilde{\Phi})^2 - \frac{\alpha}{r_0}(n+\tilde{\Phi}) - \tilde{E}^{\pm}_n(k) \right]^2} \right)^{-1}} \end{equation}
which clearly satisfy $b^{\pm}_n(-k) = b^{\pm}_n(k)$, $\left( b^{\pm}_n(k) \right)^* = b^{\pm}_n(k)$, $a^{\pm}_n(-k) = -a^{\pm}_n(k)$ and $\left( a^{\pm}_n(k) \right)^* = -a^{\pm}_n(k)$.

%
%
%
\bibliographystyle{apsrev4-1}
\bibliography{citationsAPS}

\begin{thebibliography}{52}%
\makeatletter
\providecommand \@ifxundefined [1]{%
 \@ifx{#1\undefined}
}%
\providecommand \@ifnum [1]{%
 \ifnum #1\expandafter \@firstoftwo
 \else \expandafter \@secondoftwo
 \fi
}%
\providecommand \@ifx [1]{%
 \ifx #1\expandafter \@firstoftwo
 \else \expandafter \@secondoftwo
 \fi
}%
\providecommand \natexlab [1]{#1}%
\providecommand \enquote  [1]{``#1''}%
\providecommand \bibnamefont  [1]{#1}%
\providecommand \bibfnamefont [1]{#1}%
\providecommand \citenamefont [1]{#1}%
\providecommand \href@noop [0]{\@secondoftwo}%
\providecommand \href [0]{\begingroup \@sanitize@url \@href}%
\providecommand \@href[1]{\@@startlink{#1}\@@href}%
\providecommand \@@href[1]{\endgroup#1\@@endlink}%
\providecommand \@sanitize@url [0]{\catcode `\\12\catcode `\$12\catcode
  `\&12\catcode `\#12\catcode `\^12\catcode `\_12\catcode `\%12\relax}%
\providecommand \@@startlink[1]{}%
\providecommand \@@endlink[0]{}%
\providecommand \url  [0]{\begingroup\@sanitize@url \@url }%
\providecommand \@url [1]{\endgroup\@href {#1}{\urlprefix }}%
\providecommand \urlprefix  [0]{URL }%
\providecommand \Eprint [0]{\href }%
\providecommand \doibase [0]{http://dx.doi.org/}%
\providecommand \selectlanguage [0]{\@gobble}%
\providecommand \bibinfo  [0]{\@secondoftwo}%
\providecommand \bibfield  [0]{\@secondoftwo}%
\providecommand \translation [1]{[#1]}%
\providecommand \BibitemOpen [0]{}%
\providecommand \bibitemStop [0]{}%
\providecommand \bibitemNoStop [0]{.\EOS\space}%
\providecommand \EOS [0]{\spacefactor3000\relax}%
\providecommand \BibitemShut  [1]{\csname bibitem#1\endcsname}%
\let\auto@bib@innerbib\@empty
\bibitem [{\citenamefont {Bakkers}\ and\ \citenamefont
  {Verheijen}(2003)}]{Bakkers2003}%
  \BibitemOpen
  \bibfield  {author} {\bibinfo {author} {\bibfnamefont {E.~P. A.~M.}\
  \bibnamefont {Bakkers}}\ and\ \bibinfo {author} {\bibfnamefont {M.~A.}\
  \bibnamefont {Verheijen}},\ }\href@noop {} {\bibfield  {journal} {\bibinfo
  {journal} {Journal of the American Chemical Society}\ }\textbf {\bibinfo
  {volume} {125}},\ \bibinfo {pages} {3440} (\bibinfo {year}
  {2003})}\BibitemShut {NoStop}%
\bibitem [{\citenamefont {Li}\ \emph {et~al.}(2006)\citenamefont {Li},
  \citenamefont {Qian}, \citenamefont {Xiang},\ and\ \citenamefont
  {Lieber}}]{Li2006}%
  \BibitemOpen
  \bibfield  {author} {\bibinfo {author} {\bibfnamefont {Y.}~\bibnamefont
  {Li}}, \bibinfo {author} {\bibfnamefont {F.}~\bibnamefont {Qian}}, \bibinfo
  {author} {\bibfnamefont {J.}~\bibnamefont {Xiang}}, \ and\ \bibinfo {author}
  {\bibfnamefont {C.~M.}\ \bibnamefont {Lieber}},\ }\href@noop {} {\bibfield
  {journal} {\bibinfo  {journal} {Materials Today}\ }\textbf {\bibinfo {volume}
  {9}},\ \bibinfo {pages} {18} (\bibinfo {year} {2006})}\BibitemShut {NoStop}%
\bibitem [{\citenamefont {Thelander}\ \emph {et~al.}(2006)\citenamefont
  {Thelander}, \citenamefont {Agarwal}, \citenamefont {Brongersma},
  \citenamefont {Eymery}, \citenamefont {Feiner}, \citenamefont {Forchel},
  \citenamefont {Scheffler}, \citenamefont {Riess}, \citenamefont {Ohlsson},
  \citenamefont {G\"{o}sele},\ and\ \citenamefont {Samuelson}}]{Thelander2006}%
  \BibitemOpen
  \bibfield  {author} {\bibinfo {author} {\bibfnamefont {C.}~\bibnamefont
  {Thelander}}, \bibinfo {author} {\bibfnamefont {P.}~\bibnamefont {Agarwal}},
  \bibinfo {author} {\bibfnamefont {S.}~\bibnamefont {Brongersma}}, \bibinfo
  {author} {\bibfnamefont {J.}~\bibnamefont {Eymery}}, \bibinfo {author}
  {\bibfnamefont {L.}~\bibnamefont {Feiner}}, \bibinfo {author} {\bibfnamefont
  {A.}~\bibnamefont {Forchel}}, \bibinfo {author} {\bibfnamefont
  {M.}~\bibnamefont {Scheffler}}, \bibinfo {author} {\bibfnamefont
  {W.}~\bibnamefont {Riess}}, \bibinfo {author} {\bibfnamefont
  {B.}~\bibnamefont {Ohlsson}}, \bibinfo {author} {\bibfnamefont
  {U.}~\bibnamefont {G\"{o}sele}}, \ and\ \bibinfo {author} {\bibfnamefont
  {L.}~\bibnamefont {Samuelson}},\ }\href@noop {} {\bibfield  {journal}
  {\bibinfo  {journal} {Materials Today}\ }\textbf {\bibinfo {volume} {9}},\
  \bibinfo {pages} {28} (\bibinfo {year} {2006})}\BibitemShut {NoStop}%
\bibitem [{\citenamefont {Yang}\ \emph {et~al.}(2010)\citenamefont {Yang},
  \citenamefont {Yan},\ and\ \citenamefont {Fardy}}]{Yang2010}%
  \BibitemOpen
  \bibfield  {author} {\bibinfo {author} {\bibfnamefont {P.}~\bibnamefont
  {Yang}}, \bibinfo {author} {\bibfnamefont {R.}~\bibnamefont {Yan}}, \ and\
  \bibinfo {author} {\bibfnamefont {M.}~\bibnamefont {Fardy}},\ }\href@noop {}
  {\bibfield  {journal} {\bibinfo  {journal} {Nano Letters}\ }\textbf {\bibinfo
  {volume} {10}},\ \bibinfo {pages} {1529} (\bibinfo {year}
  {2010})}\BibitemShut {NoStop}%
\bibitem [{\citenamefont {van Tilburg}\ \emph {et~al.}(2010)\citenamefont {van
  Tilburg}, \citenamefont {Algra}, \citenamefont {Immink}, \citenamefont
  {Verheijen}, \citenamefont {Bakkers},\ and\ \citenamefont
  {Kouwenhoven}}]{vanTilburg2010}%
  \BibitemOpen
  \bibfield  {author} {\bibinfo {author} {\bibfnamefont {J.~W.~W.}\
  \bibnamefont {van Tilburg}}, \bibinfo {author} {\bibfnamefont {R.~E.}\
  \bibnamefont {Algra}}, \bibinfo {author} {\bibfnamefont {W.~G.~G.}\
  \bibnamefont {Immink}}, \bibinfo {author} {\bibfnamefont {M.}~\bibnamefont
  {Verheijen}}, \bibinfo {author} {\bibfnamefont {E.~P. A.~M.}\ \bibnamefont
  {Bakkers}}, \ and\ \bibinfo {author} {\bibfnamefont {L.~P.}\ \bibnamefont
  {Kouwenhoven}},\ }\href@noop {} {\bibfield  {journal} {\bibinfo  {journal}
  {Semiconductor Science and Technology}\ }\textbf {\bibinfo {volume} {25}},\
  \bibinfo {pages} {024011} (\bibinfo {year} {2010})}\BibitemShut {NoStop}%
\bibitem [{\citenamefont {Popovitz-Biro}\ \emph {et~al.}(2011)\citenamefont
  {Popovitz-Biro}, \citenamefont {Kretinin}, \citenamefont {{Von Huth}},\ and\
  \citenamefont {Shtrikman}}]{Popovitz2011}%
  \BibitemOpen
  \bibfield  {author} {\bibinfo {author} {\bibfnamefont {R.}~\bibnamefont
  {Popovitz-Biro}}, \bibinfo {author} {\bibfnamefont {A.}~\bibnamefont
  {Kretinin}}, \bibinfo {author} {\bibfnamefont {P.}~\bibnamefont {{Von
  Huth}}}, \ and\ \bibinfo {author} {\bibfnamefont {H.}~\bibnamefont
  {Shtrikman}},\ }\href@noop {} {\bibfield  {journal} {\bibinfo  {journal}
  {Crystal Growth \& Design}\ }\textbf {\bibinfo {volume} {11}},\ \bibinfo
  {pages} {3858} (\bibinfo {year} {2011})}\BibitemShut {NoStop}%
\bibitem [{\citenamefont {Rieger}\ \emph {et~al.}(2012)\citenamefont {Rieger},
  \citenamefont {Luysberg}, \citenamefont {Sch\"{a}pers}, \citenamefont
  {Gr\"{u}tzmacher},\ and\ \citenamefont {Lepsa}}]{Rieger2012}%
  \BibitemOpen
  \bibfield  {author} {\bibinfo {author} {\bibfnamefont {T.}~\bibnamefont
  {Rieger}}, \bibinfo {author} {\bibfnamefont {M.}~\bibnamefont {Luysberg}},
  \bibinfo {author} {\bibfnamefont {T.}~\bibnamefont {Sch\"{a}pers}}, \bibinfo
  {author} {\bibfnamefont {D.}~\bibnamefont {Gr\"{u}tzmacher}}, \ and\ \bibinfo
  {author} {\bibfnamefont {M.~I.}\ \bibnamefont {Lepsa}},\ }\href@noop {}
  {\bibfield  {journal} {\bibinfo  {journal} {Nano Letters}\ }\textbf {\bibinfo
  {volume} {12}},\ \bibinfo {pages} {5559} (\bibinfo {year}
  {2012})}\BibitemShut {NoStop}%
\bibitem [{\citenamefont {Jung}\ \emph {et~al.}(2008)\citenamefont {Jung},
  \citenamefont {Lee}, \citenamefont {Song}, \citenamefont {Kim}, \citenamefont
  {Lee}, \citenamefont {Kim}, \citenamefont {Park}, \citenamefont {Choi},
  \citenamefont {Katsumoto}, \citenamefont {Lee},\ and\ \citenamefont
  {Kim}}]{Jung2008}%
  \BibitemOpen
  \bibfield  {author} {\bibinfo {author} {\bibfnamefont {M.}~\bibnamefont
  {Jung}}, \bibinfo {author} {\bibfnamefont {J.~S.}\ \bibnamefont {Lee}},
  \bibinfo {author} {\bibfnamefont {W.}~\bibnamefont {Song}}, \bibinfo {author}
  {\bibfnamefont {Y.~H.}\ \bibnamefont {Kim}}, \bibinfo {author} {\bibfnamefont
  {S.~D.}\ \bibnamefont {Lee}}, \bibinfo {author} {\bibfnamefont
  {N.}~\bibnamefont {Kim}}, \bibinfo {author} {\bibfnamefont {J.}~\bibnamefont
  {Park}}, \bibinfo {author} {\bibfnamefont {M.-S.}\ \bibnamefont {Choi}},
  \bibinfo {author} {\bibfnamefont {S.}~\bibnamefont {Katsumoto}}, \bibinfo
  {author} {\bibfnamefont {H.}~\bibnamefont {Lee}}, \ and\ \bibinfo {author}
  {\bibfnamefont {J.}~\bibnamefont {Kim}},\ }\href@noop {} {\bibfield
  {journal} {\bibinfo  {journal} {Nano Letters}\ }\textbf {\bibinfo {volume}
  {8}},\ \bibinfo {pages} {3189} (\bibinfo {year} {2008})}\BibitemShut
  {NoStop}%
\bibitem [{\citenamefont {G\"{u}l}\ \emph {et~al.}(2014)\citenamefont
  {G\"{u}l}, \citenamefont {Demarina}, \citenamefont {Bl\"{o}mers},
  \citenamefont {Rieger}, \citenamefont {L\"{u}th}, \citenamefont {Lepsa},
  \citenamefont {Gr\"{u}tzmacher},\ and\ \citenamefont
  {Sch\"{a}pers}}]{Gul2014}%
  \BibitemOpen
  \bibfield  {author} {\bibinfo {author} {\bibfnamefont {O.}~\bibnamefont
  {G\"{u}l}}, \bibinfo {author} {\bibfnamefont {N.}~\bibnamefont {Demarina}},
  \bibinfo {author} {\bibfnamefont {C.}~\bibnamefont {Bl\"{o}mers}}, \bibinfo
  {author} {\bibfnamefont {T.}~\bibnamefont {Rieger}}, \bibinfo {author}
  {\bibfnamefont {H.}~\bibnamefont {L\"{u}th}}, \bibinfo {author}
  {\bibfnamefont {M.~I.}\ \bibnamefont {Lepsa}}, \bibinfo {author}
  {\bibfnamefont {D.}~\bibnamefont {Gr\"{u}tzmacher}}, \ and\ \bibinfo {author}
  {\bibfnamefont {T.}~\bibnamefont {Sch\"{a}pers}},\ }\href {\doibase
  10.1103/PhysRevB.89.045417} {\bibfield  {journal} {\bibinfo  {journal} {Phys.
  Rev. B}\ }\textbf {\bibinfo {volume} {89}},\ \bibinfo {pages} {045417}
  (\bibinfo {year} {2014})}\BibitemShut {NoStop}%
\bibitem [{\citenamefont {Bl\"{o}mers}\ \emph {et~al.}(2013)\citenamefont
  {Bl\"{o}mers}, \citenamefont {Rieger}, \citenamefont {Zellekens},
  \citenamefont {Haas}, \citenamefont {Lepsa}, \citenamefont {Hardtdegen},
  \citenamefont {G\"{u}l}, \citenamefont {Demarina}, \citenamefont
  {Gr\"{u}tzmacher}, \citenamefont {L\"{u}th},\ and\ \citenamefont
  {Sch\"{a}pers}}]{Blomers2013}%
  \BibitemOpen
  \bibfield  {author} {\bibinfo {author} {\bibfnamefont {C.}~\bibnamefont
  {Bl\"{o}mers}}, \bibinfo {author} {\bibfnamefont {T.}~\bibnamefont {Rieger}},
  \bibinfo {author} {\bibfnamefont {P.}~\bibnamefont {Zellekens}}, \bibinfo
  {author} {\bibfnamefont {F.}~\bibnamefont {Haas}}, \bibinfo {author}
  {\bibfnamefont {M.~I.}\ \bibnamefont {Lepsa}}, \bibinfo {author}
  {\bibfnamefont {H.}~\bibnamefont {Hardtdegen}}, \bibinfo {author}
  {\bibfnamefont {O.}~\bibnamefont {G\"{u}l}}, \bibinfo {author} {\bibfnamefont
  {N.}~\bibnamefont {Demarina}}, \bibinfo {author} {\bibfnamefont
  {D.}~\bibnamefont {Gr\"{u}tzmacher}}, \bibinfo {author} {\bibfnamefont
  {H.}~\bibnamefont {L\"{u}th}}, \ and\ \bibinfo {author} {\bibfnamefont
  {T.}~\bibnamefont {Sch\"{a}pers}},\ }\href
  {http://stacks.iop.org/0957-4484/24/i=3/a=035203} {\bibfield  {journal}
  {\bibinfo  {journal} {Nanotechnology}\ }\textbf {\bibinfo {volume} {24}},\
  \bibinfo {pages} {035203} (\bibinfo {year} {2013})}\BibitemShut {NoStop}%
\bibitem [{\citenamefont {Aharonov}\ and\ \citenamefont
  {Bohm}(1959)}]{Aharonov1959}%
  \BibitemOpen
  \bibfield  {author} {\bibinfo {author} {\bibfnamefont {Y.}~\bibnamefont
  {Aharonov}}\ and\ \bibinfo {author} {\bibfnamefont {D.}~\bibnamefont
  {Bohm}},\ }\href {\doibase 10.1103/PhysRev.115.485} {\bibfield  {journal}
  {\bibinfo  {journal} {Phys. Rev.}\ }\textbf {\bibinfo {volume} {115}},\
  \bibinfo {pages} {485} (\bibinfo {year} {1959})}\BibitemShut {NoStop}%
\bibitem [{\citenamefont {Byers}\ and\ \citenamefont {Yang}(1961)}]{Byers1961}%
  \BibitemOpen
  \bibfield  {author} {\bibinfo {author} {\bibfnamefont {N.}~\bibnamefont
  {Byers}}\ and\ \bibinfo {author} {\bibfnamefont {C.~N.}\ \bibnamefont
  {Yang}},\ }\href {\doibase 10.1103/PhysRevLett.7.46} {\bibfield  {journal}
  {\bibinfo  {journal} {Phys. Rev. Lett.}\ }\textbf {\bibinfo {volume} {7}},\
  \bibinfo {pages} {46} (\bibinfo {year} {1961})}\BibitemShut {NoStop}%
\bibitem [{\citenamefont {Webb}\ \emph {et~al.}(1985)\citenamefont {Webb},
  \citenamefont {Washburn}, \citenamefont {Umbach},\ and\ \citenamefont
  {Laibowitz}}]{Webb1985}%
  \BibitemOpen
  \bibfield  {author} {\bibinfo {author} {\bibfnamefont {R.~A.}\ \bibnamefont
  {Webb}}, \bibinfo {author} {\bibfnamefont {S.}~\bibnamefont {Washburn}},
  \bibinfo {author} {\bibfnamefont {C.~P.}\ \bibnamefont {Umbach}}, \ and\
  \bibinfo {author} {\bibfnamefont {R.~B.}\ \bibnamefont {Laibowitz}},\ }\href
  {\doibase 10.1103/PhysRevLett.54.2696} {\bibfield  {journal} {\bibinfo
  {journal} {Phys. Rev. Lett.}\ }\textbf {\bibinfo {volume} {54}},\ \bibinfo
  {pages} {2696} (\bibinfo {year} {1985})}\BibitemShut {NoStop}%
\bibitem [{\citenamefont {L\'{e}vy}\ \emph {et~al.}(1990)\citenamefont
  {L\'{e}vy}, \citenamefont {Dolan}, \citenamefont {Dunsmuir},\ and\
  \citenamefont {Bouchiat}}]{Levy1990}%
  \BibitemOpen
  \bibfield  {author} {\bibinfo {author} {\bibfnamefont {L.~P.}\ \bibnamefont
  {L\'{e}vy}}, \bibinfo {author} {\bibfnamefont {G.}~\bibnamefont {Dolan}},
  \bibinfo {author} {\bibfnamefont {J.}~\bibnamefont {Dunsmuir}}, \ and\
  \bibinfo {author} {\bibfnamefont {H.}~\bibnamefont {Bouchiat}},\ }\href@noop
  {} {\bibfield  {journal} {\bibinfo  {journal} {Phys. Rev. Lett.}\ }\textbf
  {\bibinfo {volume} {64}},\ \bibinfo {pages} {2074} (\bibinfo {year}
  {1990})}\BibitemShut {NoStop}%
\bibitem [{\citenamefont {Gladilin}\ \emph {et~al.}(2013)\citenamefont
  {Gladilin}, \citenamefont {Tempere}, \citenamefont {Devreese},\ and\
  \citenamefont {Koenraad}}]{Gladilin2013}%
  \BibitemOpen
  \bibfield  {author} {\bibinfo {author} {\bibfnamefont {V.~N.}\ \bibnamefont
  {Gladilin}}, \bibinfo {author} {\bibfnamefont {J.}~\bibnamefont {Tempere}},
  \bibinfo {author} {\bibfnamefont {J.~T.}\ \bibnamefont {Devreese}}, \ and\
  \bibinfo {author} {\bibfnamefont {P.~M.}\ \bibnamefont {Koenraad}},\ }\href
  {\doibase 10.1103/PhysRevB.87.165424} {\bibfield  {journal} {\bibinfo
  {journal} {Phys. Rev. B}\ }\textbf {\bibinfo {volume} {87}},\ \bibinfo
  {pages} {165424} (\bibinfo {year} {2013})}\BibitemShut {NoStop}%
\bibitem [{\citenamefont {Tserkovnyak}\ and\ \citenamefont
  {Halperin}(2006)}]{Tserkovnyak2006}%
  \BibitemOpen
  \bibfield  {author} {\bibinfo {author} {\bibfnamefont {Y.}~\bibnamefont
  {Tserkovnyak}}\ and\ \bibinfo {author} {\bibfnamefont {B.~I.}\ \bibnamefont
  {Halperin}},\ }\href {\doibase 10.1103/PhysRevB.74.245327} {\bibfield
  {journal} {\bibinfo  {journal} {Phys. Rev. B}\ }\textbf {\bibinfo {volume}
  {74}},\ \bibinfo {pages} {245327} (\bibinfo {year} {2006})}\BibitemShut
  {NoStop}%
\bibitem [{\citenamefont {Winkler}(2003)}]{Winkler}%
  \BibitemOpen
  \bibfield  {author} {\bibinfo {author} {\bibfnamefont {R.}~\bibnamefont
  {Winkler}},\ }\href@noop {} {\emph {\bibinfo {title} {Spin orbit coupling
  effects in two-dimensional electron and hole systems}}}\ (\bibinfo
  {publisher} {Springer-Verlag},\ \bibinfo {year} {2003})\BibitemShut {NoStop}%
\bibitem [{\citenamefont {Bringer}\ and\ \citenamefont
  {Sch\"{a}pers}(2011)}]{Bringer2011}%
  \BibitemOpen
  \bibfield  {author} {\bibinfo {author} {\bibfnamefont {A.}~\bibnamefont
  {Bringer}}\ and\ \bibinfo {author} {\bibfnamefont {T.}~\bibnamefont
  {Sch\"{a}pers}},\ }\href {\doibase 10.1103/PhysRevB.83.115305} {\bibfield
  {journal} {\bibinfo  {journal} {Phys. Rev. B}\ }\textbf {\bibinfo {volume}
  {83}},\ \bibinfo {pages} {115305} (\bibinfo {year} {2011})}\BibitemShut
  {NoStop}%
\bibitem [{\citenamefont {Mehdiyev}\ \emph {et~al.}(2009)\citenamefont
  {Mehdiyev}, \citenamefont {Babayev}, \citenamefont {Cakmak},\ and\
  \citenamefont {Artunc}}]{Mehdiyev2009}%
  \BibitemOpen
  \bibfield  {author} {\bibinfo {author} {\bibfnamefont {B.}~\bibnamefont
  {Mehdiyev}}, \bibinfo {author} {\bibfnamefont {A.}~\bibnamefont {Babayev}},
  \bibinfo {author} {\bibfnamefont {S.}~\bibnamefont {Cakmak}}, \ and\ \bibinfo
  {author} {\bibfnamefont {E.}~\bibnamefont {Artunc}},\ }\href {\doibase
  10.1016/j.spmi.2009.08.009} {\bibfield  {journal} {\bibinfo  {journal}
  {Superlattices and Microstructures}\ }\textbf {\bibinfo {volume} {46}},\
  \bibinfo {pages} {593} (\bibinfo {year} {2009})}\BibitemShut {NoStop}%
\bibitem [{\citenamefont {Sheng}\ and\ \citenamefont
  {Chang}(2006)}]{Sheng2006}%
  \BibitemOpen
  \bibfield  {author} {\bibinfo {author} {\bibfnamefont {J.~S.}\ \bibnamefont
  {Sheng}}\ and\ \bibinfo {author} {\bibfnamefont {K.}~\bibnamefont {Chang}},\
  }\href {\doibase 10.1103/PhysRevB.74.235315} {\bibfield  {journal} {\bibinfo
  {journal} {Phys. Rev. B}\ }\textbf {\bibinfo {volume} {74}},\ \bibinfo
  {pages} {235315} (\bibinfo {year} {2006})}\BibitemShut {NoStop}%
\bibitem [{\citenamefont {Datta}(1995)}]{Datta}%
  \BibitemOpen
  \bibfield  {author} {\bibinfo {author} {\bibfnamefont {S.}~\bibnamefont
  {Datta}},\ }\href@noop {} {\emph {\bibinfo {title} {Electronic Transport in
  Mesoscopic Systems}}}\ (\bibinfo  {publisher} {Cambridge University Press},\
  \bibinfo {year} {1995})\BibitemShut {NoStop}%
\bibitem [{\citenamefont {Gudmundsson}\ \emph {et~al.}(2009)\citenamefont
  {Gudmundsson}, \citenamefont {Gainar}, \citenamefont {Tang}, \citenamefont
  {Moldoveanu},\ and\ \citenamefont {Manolescu}}]{Gudmundsson2009}%
  \BibitemOpen
  \bibfield  {author} {\bibinfo {author} {\bibfnamefont {V.}~\bibnamefont
  {Gudmundsson}}, \bibinfo {author} {\bibfnamefont {C.}~\bibnamefont {Gainar}},
  \bibinfo {author} {\bibfnamefont {C.-S.}\ \bibnamefont {Tang}}, \bibinfo
  {author} {\bibfnamefont {V.}~\bibnamefont {Moldoveanu}}, \ and\ \bibinfo
  {author} {\bibfnamefont {A.}~\bibnamefont {Manolescu}},\ }\href
  {http://stacks.iop.org/1367-2630/11/i=11/a=113007} {\bibfield  {journal}
  {\bibinfo  {journal} {New Journal of Physics}\ }\textbf {\bibinfo {volume}
  {11}},\ \bibinfo {pages} {113007} (\bibinfo {year} {2009})}\BibitemShut
  {NoStop}%
\bibitem [{\citenamefont {Thygesen}\ \emph {et~al.}(2003)\citenamefont
  {Thygesen}, \citenamefont {Bollinger},\ and\ \citenamefont
  {Jacobsen}}]{Thygesen2003}%
  \BibitemOpen
  \bibfield  {author} {\bibinfo {author} {\bibfnamefont {K.~S.}\ \bibnamefont
  {Thygesen}}, \bibinfo {author} {\bibfnamefont {M.~V.}\ \bibnamefont
  {Bollinger}}, \ and\ \bibinfo {author} {\bibfnamefont {K.~W.}\ \bibnamefont
  {Jacobsen}},\ }\href {\doibase 10.1103/PhysRevB.67.115404} {\bibfield
  {journal} {\bibinfo  {journal} {Phys. Rev. B}\ }\textbf {\bibinfo {volume}
  {67}},\ \bibinfo {pages} {115404} (\bibinfo {year} {2003})}\BibitemShut
  {NoStop}%
\bibitem [{\citenamefont {Kurth}\ \emph {et~al.}(2005)\citenamefont {Kurth},
  \citenamefont {Stefanucci}, \citenamefont {Almbladh}, \citenamefont {Rubio},\
  and\ \citenamefont {Gross}}]{Kurth2005}%
  \BibitemOpen
  \bibfield  {author} {\bibinfo {author} {\bibfnamefont {S.}~\bibnamefont
  {Kurth}}, \bibinfo {author} {\bibfnamefont {G.}~\bibnamefont {Stefanucci}},
  \bibinfo {author} {\bibfnamefont {C.-O.}\ \bibnamefont {Almbladh}}, \bibinfo
  {author} {\bibfnamefont {A.}~\bibnamefont {Rubio}}, \ and\ \bibinfo {author}
  {\bibfnamefont {E.~K.~U.}\ \bibnamefont {Gross}},\ }\href {\doibase
  10.1103/PhysRevB.72.035308} {\bibfield  {journal} {\bibinfo  {journal} {Phys.
  Rev. B}\ }\textbf {\bibinfo {volume} {72}},\ \bibinfo {pages} {035308}
  (\bibinfo {year} {2005})}\BibitemShut {NoStop}%
\bibitem [{\citenamefont {Brandbyge}\ \emph {et~al.}(2002)\citenamefont
  {Brandbyge}, \citenamefont {Mozos}, \citenamefont {Ordej\'{o}n},
  \citenamefont {Taylor},\ and\ \citenamefont {Stokbro}}]{Brandbygge2002}%
  \BibitemOpen
  \bibfield  {author} {\bibinfo {author} {\bibfnamefont {M.}~\bibnamefont
  {Brandbyge}}, \bibinfo {author} {\bibfnamefont {J.-L.}\ \bibnamefont
  {Mozos}}, \bibinfo {author} {\bibfnamefont {P.}~\bibnamefont {Ordej\'{o}n}},
  \bibinfo {author} {\bibfnamefont {J.}~\bibnamefont {Taylor}}, \ and\ \bibinfo
  {author} {\bibfnamefont {K.}~\bibnamefont {Stokbro}},\ }\href {\doibase
  10.1103/PhysRevB.65.165401} {\bibfield  {journal} {\bibinfo  {journal} {Phys.
  Rev. B}\ }\textbf {\bibinfo {volume} {65}},\ \bibinfo {pages} {165401}
  (\bibinfo {year} {2002})}\BibitemShut {NoStop}%
\bibitem [{\citenamefont {Tada}\ \emph {et~al.}(2004)\citenamefont {Tada},
  \citenamefont {Kondo},\ and\ \citenamefont {Yoshizawa}}]{Tada2004}%
  \BibitemOpen
  \bibfield  {author} {\bibinfo {author} {\bibfnamefont {T.}~\bibnamefont
  {Tada}}, \bibinfo {author} {\bibfnamefont {M.}~\bibnamefont {Kondo}}, \ and\
  \bibinfo {author} {\bibfnamefont {K.}~\bibnamefont {Yoshizawa}},\ }\href
  {\doibase 10.1063/1.1799991} {\bibfield  {journal} {\bibinfo  {journal} {The
  Journal of Chemical Physics}\ }\textbf {\bibinfo {volume} {121}},\ \bibinfo
  {pages} {8050} (\bibinfo {year} {2004})}\BibitemShut {NoStop}%
\bibitem [{\citenamefont {Paulsson}\ and\ \citenamefont
  {Brandbyge}(2007)}]{Paulsson2007}%
  \BibitemOpen
  \bibfield  {author} {\bibinfo {author} {\bibfnamefont {M.}~\bibnamefont
  {Paulsson}}\ and\ \bibinfo {author} {\bibfnamefont {M.}~\bibnamefont
  {Brandbyge}},\ }\href {\doibase 10.1103/PhysRevB.76.115117} {\bibfield
  {journal} {\bibinfo  {journal} {Phys. Rev. B}\ }\textbf {\bibinfo {volume}
  {76}},\ \bibinfo {pages} {115117} (\bibinfo {year} {2007})}\BibitemShut
  {NoStop}%
\bibitem [{\citenamefont {Paulsson}(2008)}]{Paulsson2008}%
  \BibitemOpen
  \bibfield  {author} {\bibinfo {author} {\bibfnamefont {M.}~\bibnamefont
  {Paulsson}},\ }\href@noop {} {} (\bibinfo {year} {2008}),\ \bibinfo {note}
  {arXiv:cond-mat/0210519v2 [cond-mat.mes-hall]}\BibitemShut {NoStop}%
\bibitem [{\citenamefont {Serra}\ and\ \citenamefont {Choi}(2009)}]{Serra2009}%
  \BibitemOpen
  \bibfield  {author} {\bibinfo {author} {\bibfnamefont {L.}~\bibnamefont
  {Serra}}\ and\ \bibinfo {author} {\bibfnamefont {M.-S.}\ \bibnamefont
  {Choi}},\ }\href@noop {} {\bibfield  {journal} {\bibinfo  {journal} {Eur.
  Phys. J. B}\ }\textbf {\bibinfo {volume} {71}},\ \bibinfo {pages} {97}
  (\bibinfo {year} {2009})}\BibitemShut {NoStop}%
\bibitem [{\citenamefont {Manolescu}\ \emph {et~al.}(2013)\citenamefont
  {Manolescu}, \citenamefont {Rosdahl}, \citenamefont {Erlingsson},
  \citenamefont {Serra},\ and\ \citenamefont {Gudmundsson}}]{Snake}%
  \BibitemOpen
  \bibfield  {author} {\bibinfo {author} {\bibfnamefont {A.}~\bibnamefont
  {Manolescu}}, \bibinfo {author} {\bibfnamefont {T.}~\bibnamefont {Rosdahl}},
  \bibinfo {author} {\bibfnamefont {S.}~\bibnamefont {Erlingsson}}, \bibinfo
  {author} {\bibfnamefont {L.}~\bibnamefont {Serra}}, \ and\ \bibinfo {author}
  {\bibfnamefont {V.}~\bibnamefont {Gudmundsson}},\ }\href {\doibase
  10.1140/epjb/e2013-40735-5} {\bibfield  {journal} {\bibinfo  {journal} {The
  European Physical Journal B}\ }\textbf {\bibinfo {volume} {86}},\ \bibinfo
  {pages} {1} (\bibinfo {year} {2013})}\BibitemShut {NoStop}%
\bibitem [{\citenamefont {Richter}\ \emph {et~al.}(2008)\citenamefont
  {Richter}, \citenamefont {Bl\"{o}mers}, \citenamefont {L\"{u}th},
  \citenamefont {Calarco}, \citenamefont {Indlekofer}, \citenamefont {Marso},\
  and\ \citenamefont {Sch\"{a}pers}}]{Richter2008}%
  \BibitemOpen
  \bibfield  {author} {\bibinfo {author} {\bibfnamefont {T.}~\bibnamefont
  {Richter}}, \bibinfo {author} {\bibfnamefont {C.}~\bibnamefont
  {Bl\"{o}mers}}, \bibinfo {author} {\bibfnamefont {H.}~\bibnamefont
  {L\"{u}th}}, \bibinfo {author} {\bibfnamefont {R.}~\bibnamefont {Calarco}},
  \bibinfo {author} {\bibfnamefont {M.}~\bibnamefont {Indlekofer}}, \bibinfo
  {author} {\bibfnamefont {M.}~\bibnamefont {Marso}}, \ and\ \bibinfo {author}
  {\bibfnamefont {T.}~\bibnamefont {Sch\"{a}pers}},\ }\href@noop {} {\bibfield
  {journal} {\bibinfo  {journal} {Nano Letters}\ }\textbf {\bibinfo {volume}
  {8}},\ \bibinfo {pages} {2834} (\bibinfo {year} {2008})}\BibitemShut
  {NoStop}%
\bibitem [{\citenamefont {Lorke}\ \emph {et~al.}(2000)\citenamefont {Lorke},
  \citenamefont {Luyken}, \citenamefont {Govorov}, \citenamefont {Kotthaus},
  \citenamefont {Garcia},\ and\ \citenamefont {Petroff}}]{Lorke2000}%
  \BibitemOpen
  \bibfield  {author} {\bibinfo {author} {\bibfnamefont {A.}~\bibnamefont
  {Lorke}}, \bibinfo {author} {\bibfnamefont {R.~J.}\ \bibnamefont {Luyken}},
  \bibinfo {author} {\bibfnamefont {A.~O.}\ \bibnamefont {Govorov}}, \bibinfo
  {author} {\bibfnamefont {J.~P.}\ \bibnamefont {Kotthaus}}, \bibinfo {author}
  {\bibfnamefont {J.~M.}\ \bibnamefont {Garcia}}, \ and\ \bibinfo {author}
  {\bibfnamefont {P.~M.}\ \bibnamefont {Petroff}},\ }\href@noop {} {\bibfield
  {journal} {\bibinfo  {journal} {Phys. Rev. Lett.}\ }\textbf {\bibinfo
  {volume} {84}},\ \bibinfo {pages} {2223} (\bibinfo {year}
  {2000})}\BibitemShut {NoStop}%
\bibitem [{\citenamefont {Takai}\ and\ \citenamefont {Ohta}(1993)}]{Takai1993}%
  \BibitemOpen
  \bibfield  {author} {\bibinfo {author} {\bibfnamefont {D.}~\bibnamefont
  {Takai}}\ and\ \bibinfo {author} {\bibfnamefont {K.}~\bibnamefont {Ohta}},\
  }\href {\doibase 10.1103/PhysRevB.48.1537} {\bibfield  {journal} {\bibinfo
  {journal} {Phys. Rev. B}\ }\textbf {\bibinfo {volume} {48}},\ \bibinfo
  {pages} {1537} (\bibinfo {year} {1993})}\BibitemShut {NoStop}%
\bibitem [{\citenamefont {Takai}\ and\ \citenamefont {Ohta}(1994)}]{Takai1994}%
  \BibitemOpen
  \bibfield  {author} {\bibinfo {author} {\bibfnamefont {D.}~\bibnamefont
  {Takai}}\ and\ \bibinfo {author} {\bibfnamefont {K.}~\bibnamefont {Ohta}},\
  }\href {\doibase 10.1103/PhysRevB.49.1844} {\bibfield  {journal} {\bibinfo
  {journal} {Phys. Rev. B}\ }\textbf {\bibinfo {volume} {49}},\ \bibinfo
  {pages} {1844} (\bibinfo {year} {1994})}\BibitemShut {NoStop}%
\bibitem [{\citenamefont {Washburn}\ \emph {et~al.}(1987)\citenamefont
  {Washburn}, \citenamefont {Schmid}, \citenamefont {Kern},\ and\ \citenamefont
  {Webb}}]{Washburn1987}%
  \BibitemOpen
  \bibfield  {author} {\bibinfo {author} {\bibfnamefont {S.}~\bibnamefont
  {Washburn}}, \bibinfo {author} {\bibfnamefont {H.}~\bibnamefont {Schmid}},
  \bibinfo {author} {\bibfnamefont {D.}~\bibnamefont {Kern}}, \ and\ \bibinfo
  {author} {\bibfnamefont {R.~A.}\ \bibnamefont {Webb}},\ }\href {\doibase
  10.1103/PhysRevLett.59.1791} {\bibfield  {journal} {\bibinfo  {journal}
  {Phys. Rev. Lett.}\ }\textbf {\bibinfo {volume} {59}},\ \bibinfo {pages}
  {1791} (\bibinfo {year} {1987})}\BibitemShut {NoStop}%
\bibitem [{\citenamefont {Nowak}\ and\ \citenamefont
  {Szafran}(2009)}]{Nowak2009}%
  \BibitemOpen
  \bibfield  {author} {\bibinfo {author} {\bibfnamefont {M.~P.}\ \bibnamefont
  {Nowak}}\ and\ \bibinfo {author} {\bibfnamefont {B.}~\bibnamefont
  {Szafran}},\ }\href {\doibase 10.1103/PhysRevB.80.195319} {\bibfield
  {journal} {\bibinfo  {journal} {Phys. Rev. B}\ }\textbf {\bibinfo {volume}
  {80}},\ \bibinfo {pages} {195319} (\bibinfo {year} {2009})}\BibitemShut
  {NoStop}%
\bibitem [{\citenamefont {Alexeev}\ and\ \citenamefont
  {Portnoi}(2012)}]{Alexeev2012}%
  \BibitemOpen
  \bibfield  {author} {\bibinfo {author} {\bibfnamefont {A.~M.}\ \bibnamefont
  {Alexeev}}\ and\ \bibinfo {author} {\bibfnamefont {M.~E.}\ \bibnamefont
  {Portnoi}},\ }\href {\doibase 10.1103/PhysRevB.85.245419} {\bibfield
  {journal} {\bibinfo  {journal} {Phys. Rev. B}\ }\textbf {\bibinfo {volume}
  {85}},\ \bibinfo {pages} {245419} (\bibinfo {year} {2012})}\BibitemShut
  {NoStop}%
\bibitem [{\citenamefont {van Oudenaarden}\ \emph {et~al.}(1998)\citenamefont
  {van Oudenaarden}, \citenamefont {Devoret}, \citenamefont {Nazarov},\ and\
  \citenamefont {Mooij}}]{Oudenaarden1998}%
  \BibitemOpen
  \bibfield  {author} {\bibinfo {author} {\bibfnamefont {A.}~\bibnamefont {van
  Oudenaarden}}, \bibinfo {author} {\bibfnamefont {M.~H.}\ \bibnamefont
  {Devoret}}, \bibinfo {author} {\bibfnamefont {Y.~V.}\ \bibnamefont
  {Nazarov}}, \ and\ \bibinfo {author} {\bibfnamefont {J.~E.}\ \bibnamefont
  {Mooij}},\ }\href@noop {} {\bibfield  {journal} {\bibinfo  {journal}
  {Nature}\ }\textbf {\bibinfo {volume} {391}},\ \bibinfo {pages} {768}
  (\bibinfo {year} {1998})}\BibitemShut {NoStop}%
\bibitem [{\citenamefont {Daday}\ \emph {et~al.}(2011)\citenamefont {Daday},
  \citenamefont {Manolescu}, \citenamefont {Marinescu},\ and\ \citenamefont
  {Gudmundsson}}]{Daday2011}%
  \BibitemOpen
  \bibfield  {author} {\bibinfo {author} {\bibfnamefont {C.}~\bibnamefont
  {Daday}}, \bibinfo {author} {\bibfnamefont {A.}~\bibnamefont {Manolescu}},
  \bibinfo {author} {\bibfnamefont {D.~C.}\ \bibnamefont {Marinescu}}, \ and\
  \bibinfo {author} {\bibfnamefont {V.}~\bibnamefont {Gudmundsson}},\ }\href
  {\doibase 10.1103/PhysRevB.84.115311} {\bibfield  {journal} {\bibinfo
  {journal} {Phys. Rev. B}\ }\textbf {\bibinfo {volume} {84}},\ \bibinfo
  {pages} {115311} (\bibinfo {year} {2011})}\BibitemShut {NoStop}%
\bibitem [{\citenamefont {Gefen}\ \emph {et~al.}(1984)\citenamefont {Gefen},
  \citenamefont {Imry},\ and\ \citenamefont {Azbel}}]{Gefen1984}%
  \BibitemOpen
  \bibfield  {author} {\bibinfo {author} {\bibfnamefont {Y.}~\bibnamefont
  {Gefen}}, \bibinfo {author} {\bibfnamefont {Y.}~\bibnamefont {Imry}}, \ and\
  \bibinfo {author} {\bibfnamefont {M.~Y.}\ \bibnamefont {Azbel}},\ }\href
  {\doibase 10.1103/PhysRevLett.52.129} {\bibfield  {journal} {\bibinfo
  {journal} {Phys. Rev. Lett.}\ }\textbf {\bibinfo {volume} {52}},\ \bibinfo
  {pages} {129} (\bibinfo {year} {1984})}\BibitemShut {NoStop}%
\bibitem [{\citenamefont {B\"{u}ttiker}\ \emph {et~al.}(1984)\citenamefont
  {B\"{u}ttiker}, \citenamefont {Imry},\ and\ \citenamefont
  {Azbel}}]{Buttiker1984}%
  \BibitemOpen
  \bibfield  {author} {\bibinfo {author} {\bibfnamefont {M.}~\bibnamefont
  {B\"{u}ttiker}}, \bibinfo {author} {\bibfnamefont {Y.}~\bibnamefont {Imry}},
  \ and\ \bibinfo {author} {\bibfnamefont {M.~Y.}\ \bibnamefont {Azbel}},\
  }\href {\doibase 10.1103/PhysRevA.30.1982} {\bibfield  {journal} {\bibinfo
  {journal} {Phys. Rev. A}\ }\textbf {\bibinfo {volume} {30}},\ \bibinfo
  {pages} {1982} (\bibinfo {year} {1984})}\BibitemShut {NoStop}%
\bibitem [{\citenamefont {Holloway}\ \emph {et~al.}(2013)\citenamefont
  {Holloway}, \citenamefont {Shiri}, \citenamefont {Haapamaki}, \citenamefont
  {Willick}, \citenamefont {Watson}, \citenamefont {LaPierre},\ and\
  \citenamefont {Baugh}}]{Holloway2013}%
  \BibitemOpen
  \bibfield  {author} {\bibinfo {author} {\bibfnamefont {G.~W.}\ \bibnamefont
  {Holloway}}, \bibinfo {author} {\bibfnamefont {D.}~\bibnamefont {Shiri}},
  \bibinfo {author} {\bibfnamefont {C.~M.}\ \bibnamefont {Haapamaki}}, \bibinfo
  {author} {\bibfnamefont {K.}~\bibnamefont {Willick}}, \bibinfo {author}
  {\bibfnamefont {G.}~\bibnamefont {Watson}}, \bibinfo {author} {\bibfnamefont
  {R.~R.}\ \bibnamefont {LaPierre}}, \ and\ \bibinfo {author} {\bibfnamefont
  {J.}~\bibnamefont {Baugh}},\ }\href@noop {} {} (\bibinfo {year} {2013}),\
  \bibinfo {note} {arXiv:1305.5552v1 [cond-mat.mes-hall]}\BibitemShut {NoStop}%
\bibitem [{\citenamefont {Ferrari}\ \emph {et~al.}(2009)\citenamefont
  {Ferrari}, \citenamefont {Goldoni}, \citenamefont {Bertoni}, \citenamefont
  {Cuoghi},\ and\ \citenamefont {Molinari}}]{Ferrari2009}%
  \BibitemOpen
  \bibfield  {author} {\bibinfo {author} {\bibfnamefont {G.}~\bibnamefont
  {Ferrari}}, \bibinfo {author} {\bibfnamefont {G.}~\bibnamefont {Goldoni}},
  \bibinfo {author} {\bibfnamefont {A.}~\bibnamefont {Bertoni}}, \bibinfo
  {author} {\bibfnamefont {G.}~\bibnamefont {Cuoghi}}, \ and\ \bibinfo {author}
  {\bibfnamefont {E.}~\bibnamefont {Molinari}},\ }\href@noop {} {\bibfield
  {journal} {\bibinfo  {journal} {Nano Letters}\ }\textbf {\bibinfo {volume}
  {9}},\ \bibinfo {pages} {1631} (\bibinfo {year} {2009})}\BibitemShut
  {NoStop}%
\bibitem [{\citenamefont {Frustaglia}\ and\ \citenamefont
  {Richter}(2004)}]{Frustaglia2004}%
  \BibitemOpen
  \bibfield  {author} {\bibinfo {author} {\bibfnamefont {D.}~\bibnamefont
  {Frustaglia}}\ and\ \bibinfo {author} {\bibfnamefont {K.}~\bibnamefont
  {Richter}},\ }\href@noop {} {\bibfield  {journal} {\bibinfo  {journal} {Phys.
  Rev. B}\ }\textbf {\bibinfo {volume} {69}},\ \bibinfo {pages} {235310}
  (\bibinfo {year} {2004})}\BibitemShut {NoStop}%
\bibitem [{\citenamefont {Nagasawa}\ \emph {et~al.}(2012)\citenamefont
  {Nagasawa}, \citenamefont {Takagi}, \citenamefont {Kunihashi}, \citenamefont
  {Kohda},\ and\ \citenamefont {Nitta}}]{Nagasawa2012}%
  \BibitemOpen
  \bibfield  {author} {\bibinfo {author} {\bibfnamefont {F.}~\bibnamefont
  {Nagasawa}}, \bibinfo {author} {\bibfnamefont {J.}~\bibnamefont {Takagi}},
  \bibinfo {author} {\bibfnamefont {Y.}~\bibnamefont {Kunihashi}}, \bibinfo
  {author} {\bibfnamefont {M.}~\bibnamefont {Kohda}}, \ and\ \bibinfo {author}
  {\bibfnamefont {J.}~\bibnamefont {Nitta}},\ }\href {\doibase
  10.1103/PhysRevLett.108.086801} {\bibfield  {journal} {\bibinfo  {journal}
  {Phys. Rev. Lett.}\ }\textbf {\bibinfo {volume} {108}},\ \bibinfo {pages}
  {086801} (\bibinfo {year} {2012})}\BibitemShut {NoStop}%
\bibitem [{\citenamefont {Sakurai}\ and\ \citenamefont
  {Napolitano}(2011)}]{Sakurai}%
  \BibitemOpen
  \bibfield  {author} {\bibinfo {author} {\bibfnamefont {J.~J.}\ \bibnamefont
  {Sakurai}}\ and\ \bibinfo {author} {\bibfnamefont {J.}~\bibnamefont
  {Napolitano}},\ }\href@noop {} {\emph {\bibinfo {title} {Modern Quantum
  Mechanics}}}\ (\bibinfo  {publisher} {Pearson},\ \bibinfo {year}
  {2011})\BibitemShut {NoStop}%
\bibitem [{\citenamefont {Cohl}\ and\ \citenamefont
  {Tohline}(1999)}]{Cohl1999}%
  \BibitemOpen
  \bibfield  {author} {\bibinfo {author} {\bibfnamefont {H.~S.}\ \bibnamefont
  {Cohl}}\ and\ \bibinfo {author} {\bibfnamefont {J.~E.}\ \bibnamefont
  {Tohline}},\ }\href@noop {} {\bibfield  {journal} {\bibinfo  {journal} {The
  Astrophysical Journal}\ }\textbf {\bibinfo {volume} {527}},\ \bibinfo {pages}
  {86} (\bibinfo {year} {1999})}\BibitemShut {NoStop}%
\bibitem [{\citenamefont {Segura}\ and\ \citenamefont
  {Gil}(2000)}]{Segura2000}%
  \BibitemOpen
  \bibfield  {author} {\bibinfo {author} {\bibfnamefont {J.}~\bibnamefont
  {Segura}}\ and\ \bibinfo {author} {\bibfnamefont {A.}~\bibnamefont {Gil}},\
  }\href {\doibase 10.1016/S0010-4655(99)00428-2} {\bibfield  {journal}
  {\bibinfo  {journal} {Computer Physics Communications}\ }\textbf {\bibinfo
  {volume} {124}},\ \bibinfo {pages} {104} (\bibinfo {year}
  {2000})}\BibitemShut {NoStop}%
\bibitem [{\citenamefont {Barticevic}\ \emph {et~al.}(2002)\citenamefont
  {Barticevic}, \citenamefont {Fuster},\ and\ \citenamefont
  {Pacheco}}]{Barticevic2002}%
  \BibitemOpen
  \bibfield  {author} {\bibinfo {author} {\bibfnamefont {Z.}~\bibnamefont
  {Barticevic}}, \bibinfo {author} {\bibfnamefont {G.}~\bibnamefont {Fuster}},
  \ and\ \bibinfo {author} {\bibfnamefont {M.}~\bibnamefont {Pacheco}},\ }\href
  {\doibase 10.1103/PhysRevB.65.193307} {\bibfield  {journal} {\bibinfo
  {journal} {Phys. Rev. B}\ }\textbf {\bibinfo {volume} {65}},\ \bibinfo
  {pages} {193307} (\bibinfo {year} {2002})}\BibitemShut {NoStop}%
\bibitem [{\citenamefont {Engels}\ \emph {et~al.}(1997)\citenamefont {Engels},
  \citenamefont {Lange}, \citenamefont {Sch\"{a}pers},\ and\ \citenamefont
  {L\"{u}th}}]{Engels1997}%
  \BibitemOpen
  \bibfield  {author} {\bibinfo {author} {\bibfnamefont {G.}~\bibnamefont
  {Engels}}, \bibinfo {author} {\bibfnamefont {J.}~\bibnamefont {Lange}},
  \bibinfo {author} {\bibfnamefont {T.}~\bibnamefont {Sch\"{a}pers}}, \ and\
  \bibinfo {author} {\bibfnamefont {H.}~\bibnamefont {L\"{u}th}},\ }\href@noop
  {} {\bibfield  {journal} {\bibinfo  {journal} {Phys. Rev. B}\ }\textbf
  {\bibinfo {volume} {55}},\ \bibinfo {pages} {R1958} (\bibinfo {year}
  {1997})}\BibitemShut {NoStop}%
\bibitem [{\citenamefont {Nitta}\ \emph {et~al.}(1997)\citenamefont {Nitta},
  \citenamefont {Akazaki}, \citenamefont {Takayanagi},\ and\ \citenamefont
  {Enoki}}]{Nitta1997}%
  \BibitemOpen
  \bibfield  {author} {\bibinfo {author} {\bibfnamefont {J.}~\bibnamefont
  {Nitta}}, \bibinfo {author} {\bibfnamefont {T.}~\bibnamefont {Akazaki}},
  \bibinfo {author} {\bibfnamefont {H.}~\bibnamefont {Takayanagi}}, \ and\
  \bibinfo {author} {\bibfnamefont {T.}~\bibnamefont {Enoki}},\ }\href
  {\doibase 10.1103/PhysRevLett.78.1335} {\bibfield  {journal} {\bibinfo
  {journal} {Phys. Rev. Lett.}\ }\textbf {\bibinfo {volume} {78}},\ \bibinfo
  {pages} {1335} (\bibinfo {year} {1997})}\BibitemShut {NoStop}%
\bibitem [{\citenamefont {Liang}\ and\ \citenamefont {Gao}(2012)}]{Liang2012}%
  \BibitemOpen
  \bibfield  {author} {\bibinfo {author} {\bibfnamefont {D.}~\bibnamefont
  {Liang}}\ and\ \bibinfo {author} {\bibfnamefont {X.~P.}\ \bibnamefont
  {Gao}},\ }\href@noop {} {\bibfield  {journal} {\bibinfo  {journal} {Nano
  Letters}\ }\textbf {\bibinfo {volume} {12}},\ \bibinfo {pages} {3263}
  (\bibinfo {year} {2012})}\BibitemShut {NoStop}%
\end{thebibliography}%

\end{document}